\documentclass[12pt,a4paper]{article}
\usepackage{graphicx,cite}

\newcommand{\figwidth}{12cm}
\newcommand{\figheight}{6.5cm}

\newcommand{\ep}{\epsilon}
\newcommand{\nn}{\nonumber}
\newcommand{\IM}{\mbox{\rm Im}}
\newcommand{\eqn}[1]{(\ref{#1})}
\newcommand{\mev}{\mbox{\rm MeV}}
\newcommand{\gev}{\mbox{\rm GeV}}
\newcommand{\Li}{\mbox{\rm Li}_2}
\newcommand{\MSb}{{\overline{\rm MS}}}
\newcommand\lsim{\mathrel{\rlap{\lower4pt\hbox{\hskip1pt$\sim$}}
    \raise1pt\hbox{$<$}}}
\newcommand\gsim{\mathrel{\rlap{\lower4pt\hbox{\hskip1pt$\sim$}}
    \raise1pt\hbox{$>$}}}

\newcommand{\jhep}[3]{{\it JHEP }{\bf #1} (#2) #3}

\newcommand{\npb}[3]{{\it Nucl. Phys. }{\bf B #1} (#2) #3}
\newcommand{\npps}[3]{{\it Nucl. Phys. }{\bf #1} {\it(Proc. Suppl.)} (#2) #3}
\newcommand{\plb}[3]{{\it Phys. Lett. }{\bf B #1} (#2) #3}

\newcommand{\prd}[3]{{\it Phys. Rev. }{\bf D #1} (#2) #3}
\newcommand{\prl}[3]{{\it Phys. Rev. Lett. }{\bf #1} (#2) #3}
\newcommand{\prep}[3]{{\it Phys. Rep. }{\bf #1} (#2) #3}
\newcommand{\rpp}[3]{{\it Rept. Prog. Phys. }{\bf #1} (#2) #3}
\newcommand{\zpc}[3]{{\it Z. Physik }{\bf C #1} (#2) #3}
\newcommand{\epjc}[3]{{\it Eur. Phys. J. }{\bf C #1} (#2) #3}
\newcommand{\sjnp}[3]{{\it Sov. J. Nucl. Phys. }{\bf #1} (#2) #3}

\newcommand{\jetpl}[3]{{\it JETP Lett. }{\bf #1} (#2) #3}
\newcommand{\ijmpa}[3]{{\it Int. J. Mod. Phys. }{\bf A #1} (#2) #3}
\newcommand{\hepph}[1]{{\tt hep-ph/#1}}

\begin{document}

\begin{titlepage}

\begin{flushright}
{\small\sf IFIC/02-31\\FTUV/02-0719} \\[15mm]
\end{flushright}

\begin{center}
{\LARGE \bf QCD moment sum rules for Coulomb systems:
 the charm and bottom quark masses}\\[10mm]

{\normalsize\bf Markus Eidem\"uller}\\[4mm]
{\small\sl Departament de F\'{\i}sica Te\`orica, IFIC,
           Universitat de Val\`encia -- CSIC,}\\
{\small\sl Apt. Correus 22085, E-46071 Val\`encia, Spain} \\[17mm]
\end{center}

\begin{abstract}
\noindent
In this work the charm and bottom quark masses are determined from QCD moment
sum rules for the charmonium and upsilon systems. To illustrate the special
character of these sum rules when applied to Coulomb systems we first set up
and study the behaviour of the sum rules in quantum mechanics. In our analysis
we include both the results from nonrelativistic QCD and perturbation theory
at next-next-to-leading order. The moments are evaluated at different values
of $q^2$ which correspond to different relative influence among the theoretical
contributions. 

In the numerical analysis we obtain the masses by choosing central values
for all input parameters. The error is estimated from a variation of these
parameters. First, the analysis is performed in the pole mass scheme.
Second, we employ the potential-subtracted mass in intermediate steps of the
calculation to then infer the quark masses in the $\MSb$-scheme. Our final
results for the pole- and $\MSb$-masses are: $M_c=1.75\pm 0.15$~GeV,
$m_c(m_c) = 1.19 \pm 0.11$~GeV, $M_b=4.98\pm 0.125$~GeV and $m_b(m_b) = 4.24
\pm 0.10$~GeV.
\end{abstract}

\vfill


\end{titlepage}

\newpage
\setcounter{page}{1}



\section{Introduction}

Quantum Chromodynamics, the fundamental theory of strong interactions, 
represents a basic building stone of the Standard Model. The determination
of its parameters remains an essential task within modern particle physics.
The strong coupling constant can be obtained from many different sources to
rather high accuracy \cite{pdg:00}. Investigations for the quark masses face
much more severe problems. Confinement effects must be taken into account for
most systems sensitive to the masses. Therefore, apart from the top mass,
nonperturbative methods like QCD sum rules \cite{svz:79,rry:85,n:89}, lattice
QCD \cite{r:92,mm:94} or chiral perturbation theory \cite{gl:85,p:95} have to
be employed.

The extraction of the heavy quark masses was among the first applications of
the method of QCD sum rules \cite{svz:79,rry:85}. In this framework the
hadronic parameters can be related to a perturbative QCD calculation including
the nonperturbative condensate contributions. The analyses were later updated
and extended \cite{dgp:94,n:94:2}. Recent times have seen a renewed
interest in these investigations. With the development of nonrelativistic QCD
(NRQCD) \cite{cl:86,bbl:95} it has been recognised that the Coulombic form
of the potential plays a major role in determinations of the charm and
bottom quark masses. The contributions from NRQCD have been calculated up to
next-next-to-leading order (NNLO) \cite{pp:99,h:99,py:98}. In a region where
the system is sensitive to mass effects, they dominate the theoretical
evaluation of the sum rules over a pure perturbative expansion in the strong
coupling constant.

The fundamental quantity in this type of sum rule analysis is the vacuum
polarisation function $\Pi(q^2)$:
\begin{equation}
  \label{eq:1.a}
  	\Pi_{\mu\nu}(q^2) = i \int d^4 x \ e^{iqx}\, \langle 
	T\{j_\mu(x) j_\nu^\dagger(0)\}\rangle
	  = (q_\mu q_\nu-g_{\mu\nu}q^2)\,\Pi(q^2)\,,
\end{equation}
where the relevant vector current is represented either by the charm 
$j_\mu^c(x)=(\bar{c}\gamma_\mu c)(x)$ or the bottom current
$j_\mu^b(x)=(\bar{b}\gamma_\mu b)(x)$.
Via the optical theorem, the experimental cross 
section $\sigma(e^+ e^- \to c\bar{c},b\bar{b})$
is related to the imaginary part of $\Pi(s)$:
\begin{equation}
  \label{eq:1.b}
	  R_{c,b}(s)=\frac{1}{Q_{c,b}^2}\,
	\frac{\sigma(e^+ e^- \to c\bar{c},b\bar{b})}
	  {\sigma(e^+ e^- \to \mu^+ \mu^-)}=12\pi\,
	\IM\, \Pi_{c,b}(s+i\ep)\,.
\end{equation}
Usually, moments of the vacuum polarisation are defined by taking derivatives
of the correlator at $s=0$. However, in this work we will allow for an
arbitrary evaluation point $s=-4 m^2 \xi$ to define the dimensionless moments:
\begin{equation}
  \label{eq:1.c}
 {\cal M}_{n}(\xi) = \frac{12\pi^2}{n!} \left(4m^2 \frac{d}{ds}\right)^n
  \Pi(s)\bigg|_{s=-4 m^2 \xi} \,.
\end{equation}
As we will see in more detail later, the parameter $\xi$ encodes much
information about the system. By taking  $\xi$ larger the evaluation
point moves further away from the threshold region. 
As already discussed in \cite{rry:85}, this leads to a better convergence
for the perturbative expansion. The price to be paid is a small dependence
of the moments on the mass. This again limits the possible accuracy when
extracting the mass from the moments.
Using a dispersion relation we can write the moments
${\cal M}_{n}(\xi)$ as an integral over the spectral density $R(s)$:
\begin{equation}
  \label{eq:1.d}
 {\cal M}_{n}(\xi) = \left(4m^2\right)^n 
 \int\limits_{s_{min}}^\infty \!ds\,\frac{R(s)}{(s+4 m^2 \xi)^{n+1}}
	= 2 \int\limits_0^1 \!dv \, \frac{v(1-v^2)^{n-1}R(v)}
	{(1+\xi(1-v^2))^{n+1}} \,.
\end{equation}
The last equation represents a convenient way to express the moments
through the velocity of the heavy quarks $v=\sqrt{1-4m^2/s}$.

The moments can either be calculated theoretically, including Coulomb
resummation, perturbation theory and nonperturbative contributions, or be
obtained from experiment. In this way one can relate the heavy quark masses
to the hadronic properties of the quark-antiquark systems. The theoretical
setup is identical for the charmonium and upsilon and in the first chapters
we will thus not specify the quark content. The difference will become crucial
for the phenomenological part and the numerical analysis. First we will
perform the analysis for the bottom quark mass where the theoretical expansions
converge better and then determine the charm quark mass from the more delicate
charmonium system.

Of decisive importance for the determination of the masses is the threshold
behaviour. Here the system reacts very sensitive to mass effects. A small
change in the mass leads to a relatively large variation of the moments. Thus
the mass can in principle be determined with rather high accuracy.
Therefore we try to develop a consistent
physical description of the threshold region which includes all theoretical
contributions and perform a matching between the low and high energy region.

Before turning to a discussion of the individual contributions, the next
section shall highlight the special character of the sum rules when applied
to the Coulomb system. To this end, we present the sum rules in the framework
of quantum mechanics which were developed in \cite{nsvz:81}. 
Then we present here for the first time the application of the moment sum rules
to a system governed by the Coulomb potential.
All theoretical
contributions can be described analytically. The sum rules show 
the dependence of the mass on the moments explicitly. Furthermore, the
dependence of the pole and continuum parts on $n$ and $\xi$ will be studied
and we investigate how these parameters determine the relative influence
between the different contributions.

The results from NRQCD can be described similar to the quantum mechanical sum
rules. They contain a nonrelativistic Green's function which is composed of a
continuous spectrum above and poles below threshold. We will isolate these
contributions and analyse their influence on the mass and the error separately.
Namely the poles will give the largest individual contribution to the mass.
Therefore we will not use the nonrelativistic expansion to the energy levels
itself which suffers from large corrections, but rather evaluate directly the
Green's function in a region where the expansion is expected to work. 

As already done in our previous work for the charmonium system \cite{ej:01},
in the theoretical description of the correlator we will use information both
from Coulomb resummation and perturbation theory. The perturbative expansion
contains a part which is not included in NRQCD and will be added to the
correlator.  The nonrelativistic spectral density is valid for low velocities
whereas the perturbative spectral density is well suited for high velocities.
Therefore we will introduce a separation velocity to construct a spectral
density for the full energy range. This will allow us to obtain a stable mass
prediction for a wide range of parameters. The matching will turn out to be
very smooth for the upsilon system but leading already to a gap for the
charmonium. The analysis will show, however, that this mismatch has no
numerical influence on the mass. The parameter $\xi$ will be used to shift
the analysis towards a more perturbative region. There the expansions of
NRQCD and perturbation theory converge faster and the different theoretical
contributions are more equally distributed. This will reduce the systematic 
uncertainty of the sum rule approach.
Since the sum rules keep the analytic dependence on the input parameters,
we will investigate in detail their influence on the mass. To estimate the
error we will vary each of the parameters in suitably large chosen windows.
The dominant uncertainty will come from the nonrelativistic expansion and
manifests itself in a large dependence on the corresponding scales.

Until now we have not specified the mass definition to be used in
eq.~\eqn{eq:1.c}. A natural choice for this mass is the pole mass $M$. In
the numerical analysis we will first use the pole mass scheme to extract the
pole masses. However, as the pole masses suffer from renormalon ambiguities
\cite{b:99}, they can only be determined up to corrections of order
$O(\Lambda_{QCD})$. Therefore, in the second part of our analysis we
shall use the potential-subtracted (PS-) mass $m_{PS}$ \cite{b:98}. From this
mass definition we can obtain the ${\rm \MSb}$-masses more accurately than
from the pole mass scheme.

In the next section we will discuss the quantum mechanical sum rules. In
section 3 we shall present the contributions from the threshold expansion in
the framework of NRQCD. Here we will also define the potential-subtracted
mass. The perturbative expansion will be derived in the following section. 
In section 5 and 6 we will discuss the nonperturbative contributions and the
phenomenological spectral function. Then we shall explain the reconstruction
of the spectral density. In the numerical analysis we will obtain the pole-
and ${\rm \MSb}$-masses from analyses in the pole- and PS-mass scheme
respectively. The origin of different contributions to the error will be
carefully investigated. After a comparison to other mass determinations we
shall conclude with a summary and an outlook.

 
\section{Quantum mechanical sum rules for the Cou\-lomb 
potential$^{\scriptscriptstyle 1}$}

\footnotetext[1]{The 
author would like to thank Matthias Jamin who has
initiated the investigations on the quantum mechanical sum rules
and has contributed a substantial part to the development of
these sum rules.}

Before studying the full quantum field theory case in detail, it will be
instructive to first investigate the corresponding quantum mechanical system.
Since it is possible to describe the system analytically, one can obtain a
clearer picture of the structure of the method and the behaviour of the
different contributions. Let us consider a system of two particles separated
by a distance $\bf x$. The Schr\"odinger equation for stationary states takes
the form
\begin{equation}
  \label{eq:2.a}
	\hat{H}\psi({\bf x})= \left[-\frac{\Delta}{2\mu}+
	V({\bf x})\right]\psi({\bf x})= E \psi({\bf x})\,,
\end{equation}
where $\mu$ represents the reduced mass of the system. The Green's function
is constructed with help of the resolvent operator
$\hat{G}(E)=(\hat{H}-E)^{-1}$.

By introducing a full set of intermediate states we obtain the
phenomenological side of the Green's function in position space:
\begin{equation}
  \label{eq:2.c}
	G({\bf x},{\bf y};E)=
	\langle{\bf x}|\hat{G}(E)|{\bf y}\rangle=
	\sum_{\alpha}\frac{\psi_\alpha({\bf x})\psi_\alpha^*({\bf y})}
	{E_\alpha-E-i\ep}+\int \, dE' \frac{\rho({\bf x},{\bf y};E')}
	{E'-E-i\ep}\,.
\end{equation}
The sum runs over the discrete part of the spectrum, $\psi_\alpha$ being the
eigenfunction to the eigenvalue $E_\alpha$. The integral is taken over the
continuum part with the spectral density $\rho({\bf x},{\bf y};E)$. By taking
the derivative at ${\bf x}={\bf y}=0$, one can define a physical correlation
function
\begin{eqnarray}
  \label{eq:2.d}
	 M(E)&=& \left[\frac{d}{dE}\,
	G({\bf x},{\bf y};E)\right]_{{\bf x}={\bf y}=0}\nn\\ 
	&=& \sum_\alpha \frac{|\psi_\alpha(0)|^2}{(E_\alpha-E-i\ep)^2 }
    	+ \int dE'\ \frac{\rho(E')}{(E'-E-i\ep)^2}\,. 
\end{eqnarray}
Via a dispersion relation, the spectral density is related to the imaginary
part of the Green's function $\rho(E)=\IM\,G(0,0;E)/\pi$. The second line of
\eqn{eq:2.d}, constituting the phenomenological part, can be compared to a
perturbative expansion of $M(E)$ thus representing the fundamental equation
for the quantum mechanical sum rules \cite{nsvz:81,bb:81,pt:84}. So far the
discussion has been general. Now we turn our attention to the Coulomb
potential:
\begin{equation}
  \label{eq:2.e}
  V(r)= -\frac{\alpha}{r} \,,\qquad E_n = -\frac{\mu\alpha^2}{2n^2}\,,\qquad 
  |\psi_{nlm}(0)|^2 = \delta_{l0}\delta_{m0} \frac{\mu^3\alpha^3}{\pi n^3}\,.
\end{equation}
In order to improve the predictive power it is convenient to formulate the
sum rules in the framework of Borel or moment sum rules. Both cases will give
valuable information: in the Borel sum rules we can relate $M(E)$ to the
retarded propagator and in this way obtain explicit analytic expressions for
the perturbative expansion. In the moment sum rules we can derive an equation
for the mass and investigate the behaviour of the pole and continuum
contributions for different values of $n$ and $\xi$.


\subsection{Borel sum rules}

In the Borel sum rules one applies a Borel transformation ${\cal B}_\tau$
to the correlation function. It is defined by
\begin{equation}
  \label{eq:2.f}
  {\cal B}_\tau = \lim_{-E,N\to \infty} \frac{(-E)^N}{\Gamma(N)}
  \frac{\partial^N}{(\partial E)^N} \,,\qquad \frac{1}{\tau}=\frac{-E}{N}
	\ \ \mbox{fixed}\,.
\end{equation}
This enhances the importance of the ground state and improves the 
perturbative expansion.
For the sum rule we will need the application of $ {\cal B}_\tau$ 
on functions of the form $(x-E)^{-\alpha}$:
\begin{equation}
  \label{eq:2.h}	
	  {\cal B}_\tau \,\frac{1}{(x-E)^\alpha} 
	= \frac{\tau^\alpha}{\Gamma(\alpha)}e^{-x\tau}\,.
\end{equation}
Now we define the Borel-transformed correlator as
\begin{eqnarray}
  \label{eq:2.i}
 	 M(\tau)&=&\frac{1}{\tau^2}\,{\cal B}_\tau \,M(E) 
	= \langle 0|e^{-\hat{H}\tau}|0\rangle \nn\\
	  &=& \sum_\alpha |\psi_\alpha(0)|^2 e^{-E_\alpha \tau} +
  	\int_0^\infty dE'\ \rho(E')e^{-E'\tau} \,.
\end{eqnarray}
Since the higher states are exponentially suppressed, the dominance
of the ground state contribution for large $\tau$ is clearly improved.
Let us now recall the definition of the retarded propagator:
\begin{equation}
  \label{eq:2.j}
  K({\bf x_f},t_f;{\bf x_i},t_i)=\langle{\bf x_f}| 
  e^{-i\hat{H}(t_f-t_i)}| {\bf x_i} \rangle \,,
\end{equation}
with $t_f>t_i$ and by analytic continuation one concludes immediately:  
\begin{equation}
  \label{eq:2.j.b}
	M(\tau) = K(0,-i\tau;0,0) \,.
\end{equation}
The known solution can be written as a product of the lowest order
perturbation theory $M_0(\tau)$ and a function $F(\gamma)$ of the dimensionless
variable $\gamma(\tau)$ \cite{v:95}:
\begin{eqnarray}
  \label{eq:2.k}
  M(\tau)&=&M_0(\tau)\,F(\gamma)\,,\qquad \gamma(\tau)=\alpha
  \sqrt{\frac{\mu\tau}{2}}\,,\quad \mbox{with} \nn\\
  M_0(\tau)&=&\left(\frac{\mu}{2\pi\tau}\right)^{3/2}\,,\nn\\
  F(\gamma)&=& 1+ 2\sqrt{\pi}\gamma+\frac{2\pi^2}{3}\gamma^2 
  +4\sqrt{\pi}\gamma^3\sum_{n=1}^\infty \frac{1}{n^3}e^{\gamma^2/n^2}
  \left(1+\mbox{erf}(\gamma/n)\right)
\end{eqnarray}
with the function ${\rm erf}(x)=(2/\sqrt{\pi})\int_0^x {\rm exp}(-t^2)\,dt$.
From this formula one can directly deduce the perturbative expansion
as a power series in small $\gamma(\tau)$:
\begin{eqnarray}
  \label{eq:2.l}
  M^{Pert}(\tau) &=& M_0(\tau)\Bigg(1+2\sqrt{\pi}\gamma+\frac{2\pi^2}{3}\gamma^2+
    4\sqrt{\pi}\gamma^3\zeta(3) \nn\\
    && +8\gamma^4\zeta(4)+4\sqrt{\pi}\gamma^5\zeta(5)
    +O(\gamma^6) \Bigg) \,.
\end{eqnarray}
On the other hand, $M(\tau)$ represents the sum of the poles and the continuum
\eqn{eq:2.i}. By inserting the energy levels and wave functions from 
eq. \eqn{eq:2.e} we can write the pole contributions as
\begin{eqnarray}
  \label{eq:2.m}
  M^{Pole}(\tau) &=& M_0(\tau)\, 8\sqrt{\pi} \gamma^3 \sum_{n=1}^{\infty}
    \frac{1}{n^3} \, e^{\gamma^2/n^2} \nn\\
	 &=& M_0(\tau)\left(8\sqrt{\pi}\gamma^3\zeta(3)+8\sqrt{\pi}
    \gamma^5\zeta(5)+O(\gamma^7)\right) \,,
\end{eqnarray}
where the second line shows the expansion for small $\gamma$.
The continuum contribution can be obtained from the difference of
eqs. \eqn{eq:2.l} and \eqn{eq:2.m}.
Let us now assume a perturbative calculation order by order in
the coupling constant. One would successively obtain the different
orders in $\gamma^n$. Up to NNLO only the continuum contribution
shows up. Then, at order $\gamma^3$, one gets a contribution
from the poles $\sim 8\sqrt{\pi}\gamma^3\zeta(3)$ which is partly cancelled
by the continuum contribution $\sim -4\sqrt{\pi}\gamma^3\zeta(3)$; 
likewise for higher orders.
The poles, starting from the order $\gamma^3$, can thus be expected
to be suppressed in the numerical evaluation.
However, this conclusion is premature.
As can be seen from eq. \eqn{eq:2.m},
the expression for $M^{Pole}$ contains an exponential in $\gamma$.
Whereas for very small $\gamma$ the poles can be safely neglected, for
large $\gamma$ these contributions grow exponentially and can far
exceed the continuum part. 

Let us finally note that the perturbative expansion is not only
an expansion in $\alpha$, but in $\gamma= \alpha\sqrt{\mu\tau/2}$
and thus depends on the Borel parameter $\tau$.
The perturbative expansion and the relative size of the poles
and the continuum will therefore depend on the sum rule window
for $\tau$ chosen in the analysis. This behaviour is better discussed
in the context of moment sum rules which closely resemble 
the field theory case at hand.


\subsection{Moment sum rules}

In the quantum mechanical Coulomb problem we define
the moments as
\begin{equation}
  \label{eq:2.n}
	M_n(\xi)=\frac{\pi}{n!}\frac{2}{\mu^2 \alpha}
	\left(\mu \alpha^2 \frac{d}{dE}\right)^n G(E)
	{\Big|}_{E=\xi E_1}\,,\quad \xi>1 \,.
\end{equation}
Again, we allow for an arbitrary evaluation point $\xi$. A natural
scale is given by the lowest bound state energy $E_1=-\mu\alpha^2/2$.
The parameter $\xi$ has been defined somewhat different as  
compared to eq. \eqn{eq:1.c}.
The derivatives must be taken in an energy region below the poles where
the Green's function is purely real. Here, we must therefore choose
$\xi>1$ whereas in eq. \eqn{eq:1.c} $\xi=0$ already represents 
a perturbative region.
Solving the relevant Schr\"odinger equation, the radial Green's function
is found to be \cite{v:95}
\begin{equation}
  \label{eq:2.o}
	G(r,0;E)=\frac{\mu k}{\pi}e^{-kr}\Gamma(1-\lambda)
	U(1-\lambda,2;2kr)
\end{equation}
with the variables
\begin{equation}
  \label{eq:2.p}
	\lambda=\frac{\mu \alpha}{k} \quad \mbox{and}\quad 
	k=\sqrt{-2\mu (E+i\ep)}\,.
\end{equation}
$U(\alpha,\beta;z)$ is the confluent hypergeometric function.
$G(r,0;E)$ is singular in the limit $r\to 0$, but the moments are
finite and the first moment $M_1(\xi)$ is found to be
\begin{equation}
  \label{eq:2.q}
	M_1(\xi)=\lambda\left(1+2\lambda+2\lambda^2\psi'(1-\lambda)\right)
	{\Big |}_{\lambda=1/\sqrt{\xi}}\,.
\end{equation}
The evaluation point $E=\xi E_1$ translates into $\lambda=1/\sqrt{\xi}$.
In this theoretical expression, all powers of $\lambda$ are resummed. 
Like the parameter $\gamma(\tau)$ in the Borel sum rules, here the true
parameter for the perturbative expansion is 
$\lambda= \alpha/\sqrt{-2E/\mu}$.
The higher moments can be derived from a recursion relation:
\begin{equation}
  \label{eq:2.r}
	M_n(\xi)=\frac{\lambda^3}{n}\frac{d}{d\lambda}
	M_{n-1}(\lambda){\Big |}_{\lambda=1/\sqrt{\xi}} \,.
\end{equation}
To derive the phenomenological parametrisation of the sum rules
we need the spectral density for positive energies which can be
obtained from
\begin{equation}
  \label{eq:2.s}
	\rho(E)=\frac{1}{\pi}\IM\, G(0,0;E)\,.
\end{equation}
$\IM \,G(r,0;E)$ is finite in the limit $r\to 0$ since it is a 
physical quantity and gives the Sommerfeld factor
\begin{equation}
  \label{eq:2.t}
	\rho(E)= \frac{\alpha \mu^2}{\pi
	\left(1-e^{-\pi \alpha\sqrt{2\mu/E}}\right)}\,, \qquad E\geq 0\,.
\end{equation}
Putting everything together, we obtain the phenomenological
parametrisation
\begin{eqnarray}
  \label{eq:2.u}
	M^{Phen}_n(\xi)&=&M^{Poles}_n(\xi)+M^{Cont}_n(\xi)
	= \sum_{k=1}^\infty
	\frac{2}{k^3 \left(\frac{E_k}{\mu\alpha^2}+\frac{\xi}{2}
	\right)^{n+1}}\nn\\
	&&+\int_0^\infty dx \, \frac{2}{\left(1-e^{-\pi\sqrt{2/x}}\right)
	\left(x+\frac{\xi}{2}\right)^{n+1}}\,.
\end{eqnarray}
Equating this quantity to the theoretical side (\ref{eq:2.q},\ref{eq:2.r})
establishes the sum rules. Using the exact formulas, this of
course represents nothing but an identity. The method comes into play
when only limited information on either part is available.
It can then, for instance, be used to extract the lowest bound state energy
by solving for $E_1$ as the higher bound states are strongly 
suppressed by the factor $1/k^3$ in eq. \eqn{eq:2.u}.
Likewise, assuming that the energy levels are known, one can
solve for the mass $\mu$ since the dominant dependence on the mass originates
from $E_1$:
\begin{equation}
  \label{eq:2.v}
	\mu=\frac{E_1}{\alpha^2}\left[\left(\frac{2}
	{M_n^{Theo}-M_n^{Cont}-M_n^{Higher Poles}}\right)^{\frac{1}{n+1}}
	-\frac{\xi}{2}	\right]^{-1}\,.
\end{equation}
Since the higher poles $M_n^{Higher Poles}$ contain $E_k$ and $\mu$
as parameters, eq. \eqn{eq:2.v} represents a fast converging
self-consistency equation for $\mu$.
Here we shall not discuss the applications of this sum rule,
but finally turn our attention to the behaviour
of the pole and continuum contributions.

To this end, we again take a closer look to eq. \eqn{eq:2.u}.
Both contributions depend in a similar way on $n$ and $\xi$.
But whereas the integration runs over positive values of $x\geq0$,
the energy levels $E_k$ are located below threshold. Let us investigate
the behaviour of the moments on $n$ for fixed $\xi$. For low values of $n$
the higher poles and high energy part of the continuum integration
can have a significant influence. When we proceed to larger $n$ we enhance
the threshold region in the continuum integration and the lowest pole
in the sum. 
Taking now $n$ fixed we see that the variation of $\xi$ can
drastically change the relative size of both parts:
Since the moments have a singularity at $\xi=1$ from $E_1$, values
of $\xi$ only slightly larger than one will lead to a complete
dominance of the first bound state on the sum rules. Larger
$\xi$ enhance the higher bound states and also the continuum part
gets more and more important. The results are summarised in 
table \ref{ta:2.a}.
\begin{table}
\begin{center}
\begin{tabular}{|c||c|c|c|c||c|c|c|c|}\hline
& \multicolumn{4}{c||}{$\xi=4$} & \multicolumn{4}{c|}{$\xi=10$} \\ \hline 
$n$ & 1 & 3 & 5 & 7 & 1 & 3 & 5 & 7  \\ \hline 
$M_n^{Poles}/M_n^{Theo}$ & 0.45 & 0.83 & 0.94 & 0.97  
& 0.16 & 0.49 & 0.67 & 0.78  \\ 
$M_n^{Cont}/M_n^{Theo}$  & 0.55 & 0.17 & 0.06 & 0.03 & 0.84 
& 0.51 & 0.33 & 0.22   \\ \hline
\hline
& \multicolumn{4}{c||}{$n=3$} & \multicolumn{4}{c|}{$n=7$} \\ \hline 
$\xi$ & 2 & 4 & 10 & 50 & 2 & 4 & 10 & 50  \\ \hline 
$M_n^{Poles}/M_n^{Theo}$ & 0.98 & 0.83 & 0.49 & 0.10  & 0.9998 
& 0.97 & 0.78 & 0.26  \\ 
$M_n^{Cont}/M_n^{Theo}$  & 0.02 & 0.17 & 0.51 & 0.90 & 0.0002 
& 0.03 & 0.22 & 0.74   \\ \hline
\end{tabular}
\caption{\label{ta:2.a}Relative size of pole and continuum contributions
to the theoretical moments $M_n^{Theo}=M_n^{Poles}+M_n^{Cont}$
for different values of $n$ at fixed $\xi$ and for different $\xi$
at fixed $n$.}
\end{center}
\end{table}
We have depicted the pole and continuum contributions for different
values of $n$ and $\xi$. One should keep in mind that the relative
size of the pole and continuum contributions does not directly
depend on the physical system under investigation 
but rather on the values of $n$ and $\xi$ chosen
for the analysis; namely the poles can depend strongly on these 
parameters. However, it is important to note that the main mass 
dependence originates from the first bound states since the continuum
is largely independent of the mass. This remains true even in a region
where the continuum dominates the moments. Therefore, 
to obtain good accuracy when extracting e.g. the
ground state energy or the mass it is advantageous to use
low $\xi$ and high $n$. Then the contribution from $E_1$ will 
dominate the sum rules. Unfortunately, in this region also the perturbative
expansion converges more slowly. The perturbative series behaves better
for lower $n$ and also for higher $\xi$ since this parameter enters
directly in the expansion variable $\lambda=1/\sqrt{\xi}$. 
Therefore, in practical applications where the exact solutions
are not known, one must carefully choose a range of values
for $n$ and $\xi$ such that the theoretical calculation can be
reliably trusted without loosing sensitivity on the parameters
one would like to extract. These considerations will be made more
explicit in the numerical analysis. Now we will discuss the Coulomb
contributions in the full field theory system.

 
\section{Coulomb resummation}

The theory of NRQCD provides a consistent framework
to treat the problem of heavy quark-antiquark production close to
threshold. 
The contributions can be described by a nonrelativistic Schr\"odinger
equation and systematically calculated in time-independent perturbation
theory (TIPT) \cite{ps:91}. At NNLO the factorised formulation has first
been shown in \cite{ht:98}.  
The correlator is expressed in terms of a Green's function 
$G(k)=G(0,0,k)$:
\begin{equation}
  \label{eq:3.a}
  \Pi(s)=\frac{N_c}{2M^2}\left(C_h(\alpha_s)G(k)+\frac{4k^2}{3M^2}G_C(k)\right)\,,
\end{equation}
where $N_c$ is the number of colours, $k=\sqrt{M^2-s/4}$ and 
$M$ represents the pole mass \cite{pp:99}.
First we will present the method in the pole mass scheme and afterwards
discuss the PS-scheme.
The constant $C_h(\alpha_s)$ is a perturbative coefficient
needed for the matching between the full and the nonrelativistic 
theory. It naturally depends on the hard scale and is given by \cite{ht:98}
\begin{eqnarray}
  \label{eq:3.b}
  C_h(\alpha_s) &=& 1-4C_F \,\frac{\alpha_s(\mu_{hard})}{\pi}
+C_h^{(2)} C_F \,\left(\frac{\alpha_s(\mu_{hard})}{\pi}\right)^2 \,,\nn\\
  C_h^{(2)} &=& \left(\frac{39}{4}-\zeta(3)+\frac{4\pi^2}{3}\ln 2-\frac{35\pi^2}{18}
    \right)C_F - \Bigg(\frac{151}{36}+\frac{13}{2}\zeta(3)\nn\\
      &&+\frac{8\pi^2}{3}\ln 2
      -\frac{179\pi^2}{72}\Bigg)C_A 
    +\left(\frac{44}{9}-\frac{4\pi^2}{9}+
      \frac{11}{9}n_f\right)T \nn\\
    &&+2 b_0 \ln\left(\frac{M}{\mu_{hard}}\right)
    +\pi^2\left(\frac{2}{3}C_F+C_A\right)\ln\left(\frac{M}{\mu_{fac}}\right) \,,
\end{eqnarray}
where $C_F=4/3,\ C_A=3,\ T=1/2$ and $b_0=11-2n_f/3$.
$G_C(k)$ represents the Coulomb Green's function and reads
\begin{equation}
  \label{eq:3.c}
  G_C(k)=-\frac{C_F \alpha_s M^2}{4 \pi}\left[\frac{k }{C_F \alpha_s M}
    +\ln\left(\frac{k}{\mu_{fac}}\right)+\gamma_E
    +\Psi\left(1-\frac{C_F \alpha_s M}{2k}\right) \right] \,.
\end{equation}
The contributions from NRQCD are summarised in the potential.
The Green's function obeys the corresponding Schr\"odinger equation
\begin{equation}
  \label{eq:3.d}
  \Bigg( -\frac{\Delta_x}{M}+V_C(x)+\Delta V(x)
  +\frac{k^2}{M}\Bigg)
  G({\bf x},{\bf y},k)=\delta^{(3)}({\bf x}-{\bf y}) \,.
\end{equation}
Here $V_C(x)=-C_F\alpha_s/|{\bf x}|$ represents the Coulomb potential
and $\Delta V(x)$ contains the NLO and NNLO corrections.
The explicit form of the potential is given in the appendix.
The full Green's function can be derived from TIPT and to first 
order one obtains
\begin{eqnarray}
  \label{eq:3.e}
  G(0,0,k) &=& G_C(0,0,k)+\Delta G(0,0,k) \,,\nn\\
  \Delta G(0,0,k) &=& -\int d^3{\bf x}\ G_C(0,{\bf x},k)
	\Delta V({x})
  G_C({\bf x},0,k) \,.
\end{eqnarray}
To be consistent to NNLO one also has to apply second order
perturbation theory to the one-loop potential. Details about this
procedure can be found in \cite{pp:99,h:99}.
To calculate the moments from the Green's function we will directly
perform the derivatives at $s=-4 M^2\xi$ according to eq. \eqn{eq:1.c}.
Since the Green's function is known analytically \cite{pp:99}
as a function of $k=k(s)$, this can be done numerically. In this way
we take advantage of the fact that the perturbative expansion parameter
depends on the evaluation point. The expansion of the moments shows
the same behaviour as has already been discussed in the quantum mechanical sum
rules (\ref{eq:2.q},\ref{eq:2.r}). There the expansion parameter of the moments
is $\lambda=1/\sqrt{\xi}$ and so higher values for $\xi$ improve the
perturbative series. The resulting moments include both the pole and the
continuum contributions.

The moments depend on three scales: 
The hard scale $\mu_{hard} \sim M$ enters in the coefficient $C_h$ 
of eq. \eqn{eq:3.a}. This scale is also needed for the perturbative
expansion which will be discussed in the next section.
The soft scale $\mu_{soft}\sim M v$ is a typical scale for nonrelativistic
processes and the relevant scale for the expansion of the Green's
function. Furthermore, the factorisation scale
$\mu_{fac}$ separates the contributions of large and small 
momenta and plays the role of an infrared cutoff.
As we perform the calculation only up to NNLO we are left with a 
residual dependence on these three scales. In fact, the dependence
of the mass on the scales, especially on $\mu_{soft}$, is rather
large and will give the dominant source of the error.
To obtain the central values for the masses we will use a set of
values for  $\mu_{soft}, \mu_{fac}$ and $\mu_{hard}$ according
to the physical expectations from the charmonium and bottomium systems.
The error will then be estimated by allowing for sufficiently large
variations of these scales.

Though the full theoretical moments from resummation can thus be 
determined, we are also interested in the pole and continuum contributions
separately. First, we want to analyse them independently and estimate
their contribution to the error. Second, in our
numerical analysis we will reconstruct the spectral density above 
threshold. At low velocities it is given by the imaginary
part of the nonrelativistic Green's function.

In principle, the expressions for the energies and decay widths of the
poles have been calculated at NNLO.
One could then deduce their contribution to the moments as in
eq. \eqn{eq:2.u}. But in this method the contributions have to
be calculated near threshold and thus show large corrections
already for the bottomium and cannot be trusted for the
charmonium.

Therefore we will choose a different
method of evaluation \cite{ej:01,ej:01:2}. 
By using a dispersion relation, we derive
the continuum from the imaginary part of the correlator. 
From the difference we can then obtain the pole contributions:
\begin{eqnarray}
  \label{eq:3.f}
  {\cal M}_n^{Poles} &=& \frac{12 \pi^2}{n!}\left(4M^2\frac{d}{ds}\right)^n
    \Pi(s)\Big|_{s=-4M^2 \xi}\nn\\
    &&-12 \pi \left(4M^2\right)^n \int_{4M^2}^\infty ds\  
    \frac{\IM \,\Pi(s)}{(s+4M^2\xi)^{n+1}}\,.
\end{eqnarray}
Nevertheless, for values of $n$ and $\xi$ used in our analysis,
the poles will give the largest contribution
to the theoretical moments and thus the dependence on the
scales will remain relatively strong. 
In the numerical analysis we will give a detailed account on
the size and behaviour of these contributions.

\begin{figure}
\begin{center}
\includegraphics[height=\figwidth,width=\figheight,angle=-90]{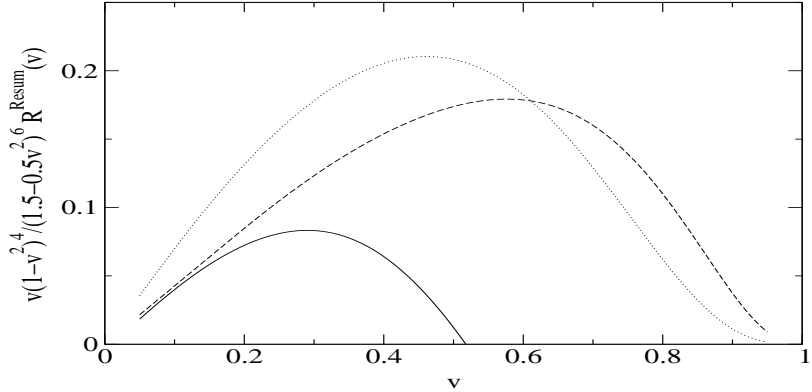}
\caption{\label{fig:3.a}
Resummed spectral density times the weight factor corresponding to
$n=5$ and $\xi=0.5$ in LO (dotted), NLO (dashed) and
NNLO (solid) at typical scales for the charmonium 
$\mu_{soft}=1.1$ GeV, $\mu_{fac}=1.45$ GeV and $\mu_{hard}=1.75$ GeV.}
\end{center}
\end{figure}

Now we investigate the spectral density from the continuum part.
We discuss the charmonium system since the differences in the
expansion can be seen more clearly than in the bottomium
which shows a faster convergence. In figure \ref{fig:3.a} we have
displayed the spectral density times the weight factor for the
different orders. The area under the spectral density is  
directly proportional to the moments. We have chosen moments and
scales typical for the numerical analysis, $n=5$, $\xi=0.5$,
$\mu_{soft}=1.1$ GeV, $\mu_{fac}=1.45$ GeV and $\mu_{hard}=1.75$ GeV.
The dotted line represents the LO, the dashed line includes the NLO
and the solid line represents the full NNLO result.
One can see that the expansion of the moments converges well in the 
low velocity region.
But when higher $v$ are used, resummation is not capable to incorporate
the correct high energy behaviour. Therefore, when we reconstruct
the complete spectral density, we will use the resummed spectral 
density only below a separation velocity $v<v_{sep}$ where the 
expansion can be trusted.

Part of the large corrections to the potential
is not inherent to the bound state system
but to the definition of the pole mass.
In \cite{hssw:99,b:98} it was observed that the long distance sensitivity
in the coordinate space potential cancels to all orders in perturbation
theory with the long distance sensitivity in the pole mass.
Therefore a new mass definition has been proposed, the 
potential-subtracted mass $m_{PS}$, where the potential below a
separation scale $\mu_{sep}$ is subtracted:
\begin{equation}
  \label{eq:3.g} 
  m_{PS}(\mu_{sep}) = M-\delta  m(\mu_{sep})\,,\quad
  \delta m(\mu_{sep}) = -\frac{1}{2}\int\limits_{|{\bf q}|<\mu_{sep}}
  \!\!\!\frac{d^3 q}{(2\pi)^3}\,V(q)\,.
\end{equation}
The subtracted potential $V(r,\mu_{sep})$ is then defined by
\begin{equation}
  \label{eq:3.h} 
  V(r,\mu_{sep}) = V(r)+2\delta m(\mu_{sep})\,.
\end{equation}
When using these definitions the Schr\"odinger equation for the 
Green's function takes it usual form where the pole mass is substituted
by the PS-mass and the potential by $V(r,\mu_{sep})$ \cite{b:98}.
Since the renormalon contributions have been subtracted from the
potential, the convergence of the expansion is improved and the strength
of the potential is reduced. The developed methods for the evaluation
of the Green's function can again be employed and the additional
contribution is absorbed in a shift of the energy.

We then express all moments in terms of the PS-mass and perform
the analysis for $m_{PS}$. The PS-mass is perturbatively related
to the ${\rm \MSb}$-mass:
\begin{eqnarray}
  \label{eq:3.i}
  m_{PS}(\mu_{sep}) &=& m \Bigg[ 1+ \frac{\alpha_s(m)}{\pi}
  \left(k_1-C_F\frac{\mu_{sep}}{m}\right)
  + \left(\frac{\alpha_s(m)}{\pi}\right)^2 \Bigg(k_2  -
	C_F\frac{\mu_{sep}}{m}\nn\\&&\times
	\frac{w_1(m,\mu_{sep})}{4} \Bigg) 
    + \left(\frac{\alpha_s(m)}{\pi}\right)^3\left(k_3
	-C_F\frac{\mu_{sep}}{m}\frac{w_2(m,\mu_{sep})}{16}\right)
      \Bigg] \,,\nn\\
  k_1 &=& C_F \,, \quad k_2 = 13.443 - 1.041 n_f \,,\nn\\
  k_3 &=& 190.595 -26.655 n_f + 0.653 n_f^2 \,,
\end{eqnarray}
where $m=m_{\MSb}(m_{\MSb})$ is the ${\rm \MSb}$-mass evaluated at its own
scale. The functions $w_1$ and $w_2$ can be found in the appendix where
also a more complete list of formulas to the PS-mass is given.

The definition of the PS-mass and its relation to the  ${\rm \MSb}$-mass
depends on $\mu_{sep}$. This scale must be taken large enough to
guarantee a perturbative relation between the masses. 
At the same time it should be chosen smaller than a typical nonrelativistic 
scale as not to affect the threshold behaviour:
\begin{equation}
  \label{eq:3.j}
  \Lambda_{QCD}<\mu_{sep}<M\cdot v.
\end{equation}
In the numerical analysis we will see that the use of the PS-mass
improves the error of the  ${\rm \MSb}$-mass. To estimate the
error on $\mu_{sep}$, we shall also vary this scale in appropriate ranges.


\section{Perturbative expansion}

The perturbative spectral function $R^{Pert}(s)$ can be expanded in 
powers of the strong coupling constant $a=\alpha_s/\pi$,
\begin{equation}
  \label{eq:4.a}
  R^{Pert}(s)= R^{(0)}(s)+ a\,R^{(1)}(s)+ 
  a^2 \,R^{(2)}(s)+O(a^3)\,.
\end{equation}
From this expression the corresponding moments
${\cal M}_n^{Pert}(\xi)$ can be calculated via the integral of eq. \eqn{eq:1.d}.
For the first two terms, analytic expressions are available 
\cite{jp:97} and they read 
\begin{eqnarray}
  \label{eq:4.b}
  R^{(0)}(v) &=& \frac{3}{2}v(3-v^2) \,,\nn\\
  R^{(1)}(v) &=& 2(1+v^2)(3-v^2)[ 4\Li(p)+2\Li(-p)+\ln(p)(\ln(1+p)\nn\\
    && +2\ln(1-p))]-4v(3-v^2)(\ln(1+p)+2\ln(1-p))\nn\\
  && -\frac{1}{4}(1-v)(33-39v-17v^2+7v^3)\ln(p)+\frac{3}{2}v(5-3v^2) \,,
\end{eqnarray}
where $p=(1-v)/(1+v)$ and $\Li$(z) is the dilogarithmic function.
The corresponding formulas for $\Pi^{(0)}$ and $\Pi^{(1)}$ can,
for instance, be found in \cite{g:84,bft:93}.
$R^{(2)}(s)$ is still not fully known
analytically. We employ a method based on  
Pad{\'e}-approximants to construct the spectral density in the
full energy range \cite{cks:96,cks:97}. It uses available
information around $q^2=0$, at threshold and in the high
energy region. 
It has the advantage that it gives a good
description until relatively close to threshold.
In this region the moments show a
strong variation for relatively small changes of the mass. A pure high energy
expansion would only be valid for large values of the velocity and a matching
between the threshold and the perturbative region would be less reliable.

\begin{figure}
\begin{center}
\includegraphics[height=\figwidth,width=\figheight,angle=-90]{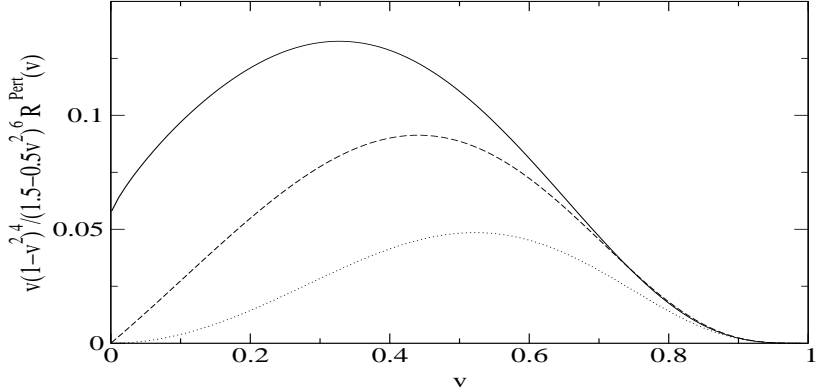}
\caption{\label{fig:4.a}
Perturbative spectral density times the weight factor corresponding to
$n=5$ and $\xi=0.5$ in LO (dotted), NLO (dashed) and
NNLO (solid) for $\mu_{hard}=1.75$ GeV.}
\end{center}
\end{figure}
To illustrate the perturbative convergence we compare 
the different orders for the charmonium as in the last section
for the resummed spectral density.
In figure \ref{fig:4.a} we have displayed
the spectral density times the weight factor, again for values of
$n=5$, $\xi=0.5$ and $\mu_{hard}=1.75$ GeV.
The expansion converges well in the
high velocity region. As we approach lower $v$, the expansion cannot
be trusted since singular terms in $v$ appear which have to be
resummed. These are included in the resummed spectral density
which sums up terms of order $\alpha_s^n/v^{n-k}$,
for $n\geq 0$ and $k=1,2,3$. The leading term in the perturbative
spectral density at NNLO has a singular behaviour $\sim \alpha_s^2/v$,
but its contribution to the moments remains finite since the weight
function contains a factor of $v$. Consequently, the graph at NNLO
in figure \ref{fig:4.a} starts with a constant set-off at $v=0$.

As will be explained in more detail in section 7, for the perturbative 
moments we will therefore mainly use the spectral density above a 
separation velocity $v>v_{sep}$ with $v_{sep}\approx 0.4$. 
\begin{table}
\begin{center}
\begin{tabular}{|c|c||c|c|c|c|}\hline
\multicolumn{6}{|c|}{$\xi=0$} \\ \hline 
\multicolumn{2}{|c||}{$v_{sep}$} & 0 & 0.2 & 0.4 & 0.6 \\ \hline 
& LO & 0.31 & 0.29 & 0.18 & 0.055  \\ 
${\cal M}_5^{Pert}$ & NLO & 0.72 & 0.60 & 0.32 & 0.081 \\ 
& NNLO& 1.29 & 0.87 & 0.38 & 0.085 \\ \hline\hline
\multicolumn{6}{|c|}{$\xi=0.5$} \\ \hline 
\multicolumn{2}{|c||}{$v_{sep}$} & 0 & 0.2 & 0.4 & 0.6 \\ \hline 
& LO & 0.046 & 0.044 & 0.033 & 0.014  \\ 
${\cal M}_5^{Pert}$ & NLO & 0.096 & 0.085 & 0.054 & 0.020  \\ 
& NNLO& 0.15  & 0.12 & 0.064 & 0.021   \\ \hline\hline
\multicolumn{6}{|c|}{$\xi=1$} \\ \hline 
\multicolumn{2}{|c||}{$v_{sep}$} & 0 & 0.2 & 0.4 & 0.6 \\ \hline 
& LO & 0.012 & 0.011 & 0.0092 & 0.0048  \\ 
${\cal M}_5^{Pert}$ & NLO & 0.023 & 0.021 & 0.015 & 0.0068  \\ 
& NNLO& 0.034 & 0.027 & 0.017 & 0.0070 \\ \hline\hline
\multicolumn{6}{|c|}{$\xi=2$} \\ \hline 
\multicolumn{2}{|c||}{$v_{sep}$} & 0 & 0.2 & 0.4 & 0.6 \\ \hline 
& LO & 0.0017 & 0.0017 & 0.0015 & 0.00093  \\ 
${\cal M}_5^{Pert}$ & NLO & 0.0030 & 0.0028 & 0.0023 & 0.0013  \\ 
& NNLO& 0.0041  & 0.0035 & 0.0025 & 0.0013   \\ \hline
\end{tabular}
\caption{\label{tab:4.a}Perturbative moments at LO, NLO and NNLO
with $\mu_{hard}=1.75$ GeV and $n=5$ for different values of $\xi$.
The moments are calculated only from the perturbative spectral
density above $v>v_{sep}$.}
\end{center}
\end{table}
In table \ref{tab:4.a} we compare the behaviour
of the moments for different values of $\xi$ and $v_{sep}$.
The higher $v_{sep}$ and $\xi$ one chooses the more one approaches 
the perturbative region and the expansion improves. 
For typical values of the analysis, $v_{sep}=0.4$ and $\xi=0.5$, 
the convergence is under good control.

To calculate the moments in the PS-scheme we can use the same 
integration formula \eqn{eq:1.d} as in the pole mass scheme,
but now the spectral density $R(s)$ is evaluated at the velocity
$v=\sqrt{1-4m_{PS}^2/s}$ and the start of the integration
$v_{sep}$ must be transformed to this scheme as well.

\begin{figure}
\begin{center}
\includegraphics[width=4cm,height=4cm]{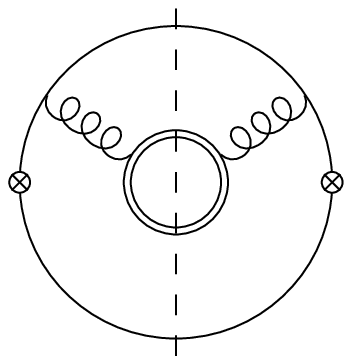}
\hspace*{2cm}
\includegraphics[width=4cm,height=4cm]{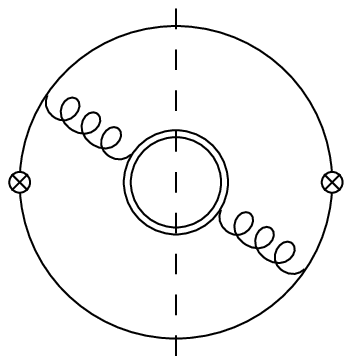}
\caption{\label{fig:4.b} Diagrams at $O(\alpha_s^2)$ from the 
light quark correlator contributing to heavy quark production.}
\end{center}
\end{figure}
At $O(\alpha_s^2)$ it is no longer true that heavy quark production
originates exclusively from the heavy quark correlator. Also the
light quark correlator includes a four fermion cut with a heavy
quark pair radiated off the light quarks \cite{pr:02}. The corresponding
diagrams are shown in fig. \ref{fig:4.b}.
When these contributions are included in the measurements, which 
of course depends on the experimental setup, they should be considered
in the theoretical side as well. The diagrams have been calculated
in ref. \cite{hjkt:94}. The resulting expressions can be split in
such a way that they allow to introduce Coulomb resummation
effects of the heavy quark pair in a straightforward way. However, 
as was discussed in \cite{pr:02}, the heavy quark pair is produced
in a colour octet state from the gluon splitting. In this case
the potential becomes repulsive and the cross section decreases
close to threshold. For high energies the diagram gives the same contribution
as the diagram with the light and heavy quark lines interchanged.
Since the main contribution of this diagram comes from the perturbative
region, its $O(\alpha_s^2)$ contribution to the moments is
suppressed and has a typical relative size of $2\cdot 10^{-5}$.
The shift in the final value for the ${\rm \MSb}$-bottom quark mass then amounts
to $\Delta m_b(m_b)\approx 9$ keV and can be safely neglected within
the uncertainty of this analysis.
To conclude the theoretical side of the correlator we now 
discuss the condensate contributions.


\section{Condensate contributions}

The nonperturbative effects on the vacuum correlator are parametrised
by the condensates. The leading term is the gluon condensate
contribution. It has been calculated up to next-to-leading order 
\cite{bbifts:94} and can be written as follows:
\begin{equation}
  \label{eq:5.a}
  \Pi^{Cond}_{FF}(s) = \frac{\langle a FF\rangle}{16M^4}
	\left(C_{FF}^{(0)}(s) +a\,C_{FF}^{(1)}(s) +O(a^2)\right) \,.
\end{equation}
The analytic form of the functions $C_{FF}^{(0)}(s)$ and $C_{FF}^{(1)}(s)$
can be found in \cite{bbifts:94}. The results have been calculated
in dimensional regularisation with the pole mass $M$. Using a
different mass prescription like the PS-mass,  $C_{FF}^{(1)}(s)$
has to change accordingly.
In our analysis below, we employ a value of 
$\langle a FF\rangle = 0.024\pm 0.012 \ \gev^4$ 
for the gluon condensate.

Furthermore, in \cite{nr:83:1,nr:83:2,bg:85} the dimension 6 and 8
condensate contributions have been calculated. 
However, this has been done only 
for moments at $\xi=0$ and therefore only in this case we take them
into account. For typical values of charmonium scales, 
$\mu_{hard}=M=1.75$ GeV, their contribution is 5\%-10\% of the leading
gluon condensate. In fact, as will be shown in the numerical
analysis, the absolute contribution of the condensates
to the full theoretical moments is small, both for the
upsilon and the charmonium. Whereas former
sum rule analyses for the charmonium have emphasised the significance
of these nonperturbative contributions, their relative suppression
in this work is due to three reasons: First, the absolute value of the
theoretical moments increases from the Coulomb resummation. 
Then we evaluate the moments at larger $\xi$ and smaller $n$ 
where the nonperturbative 
contributions are relatively small. Finally, since we obtain a larger 
pole mass than former analyses, the condensates,
starting with a power of $1/M^4$, are suppressed further.


\section{Phenomenological spectral function}

Experimentally, the six lowest lying $\psi-$ and $\Upsilon-$resonances 
have been observed.
To obtain the phenomenological content of the spectral function we
use the narrow-width  approximation for the resonances
\begin{equation}
  \label{eq:6.a}
  R_k(s) = \frac{9\pi}{{\overline \alpha}^2 Q_{c,b}^2}\,
  \Gamma_k(\psi_k,\Upsilon_k \to e^+e^-)\,E_k\,\delta(s-E_k^2)\,,
\end{equation}
where $Q_{c,b}$ represents the electric charge of the charm or
bottom quark. ${\overline \alpha}$ denotes the running QED coupling
evaluated at a scale around the resonance mass.
For the charm this corresponds to the fine structure constant
$\alpha=1/137.04$ whereas for the bottom widths the Review of
Particle Properties \cite{pdg:00} has used ${\overline \alpha}^2
= 1.07 \alpha^2$ and we will do so accordingly. The 
narrow-width approximation provides an excellent description
of these states since the full hadronic widths are much smaller 
than the masses. 
The values for the masses and electronic widths are collected in
tables \ref{tab:6.a} and \ref{tab:6.b}.
For our numerical analysis the errors on the masses can be safely
neglected and have thus not been listed.
It should be kept in mind that the moments from the experimental
resonances are not identical to the ones obtained from the poles
of the Green's function in section 3 which represent a summation
of a special kind of theoretical contributions.
\begin{table}
\begin{center}
\begin{tabular}{|c||c|c|c|}\hline
$k$  & 1 & 2 & 3  \\ \hline\hline
$E_k$ [GeV] & 3.097 & 3.686 & 3.770 \\ \hline
$\Gamma_k$ [keV] & $5.26\pm 0.37$ & $2.12\pm 0.18$ & $0.24\pm 0.05$  \\ \hline\hline
$k$  & 4 & 5 & 6 \\ \hline\hline
$E_k$ [GeV] & $4.040$ & $4.159$ & $4.415$ \\ \hline
$\Gamma_k$ [keV] & $0.75\pm 0.15$
 & $0.77\pm 0.23$ & $0.47\pm 0.10$ \\ \hline
\end{tabular}
\caption{\label{tab:6.a}Masses and electronic widths of the
first six $\psi_k$-resonances.}
\end{center}
\end{table}
\begin{table}
\begin{center}
\begin{tabular}{|c||c|c|c|}\hline
$k$  & 1 & 2 & 3  \\ \hline\hline
$E_k$ [GeV] & 9.460 & 10.023 & 10.355 \\ \hline
$\Gamma_k$ [keV] & $1.32\pm 0.07$ & $0.52\pm 0.04$ & 
$0.48\pm 0.08$  \\ \hline\hline
$k$  & 4 & 5 & 6 \\ \hline\hline
$E_k$ [GeV] & 10.580 & 10.865 & 11.019 \\ \hline
$\Gamma_k$ [keV] & $0.25\pm 0.03$
 & $0.31\pm 0.07$ & $0.13\pm 0.03$ \\ \hline
\end{tabular}
\caption{\label{tab:6.b}Masses and electronic widths of the
first six $\Upsilon_k$-resonances.}
\end{center}
\end{table}

For the upsilon system, the hadronic continuum is not measured with
sufficient accuracy so we use the assumption of quark-hadron duality 
and integrate the theoretical spectral density above a continuum
threshold $s_0$:
\begin{equation}
  \label{eq:6.b}
  \frac{{\cal M}_{b,n}}{(4M_b^2)^n} = \frac{9\pi}{{\overline \alpha}^2 
	Q_{b}^2}\sum_{k=1}^6 \frac{\Gamma_{b,k} \, E_{b,k}}
	{(E_{b,k}^2+4M_b^2\xi)^{n+1}}
  + \int_{s_0}^\infty ds\,\frac{R^{Rcstr}_b(s)}{(s+4M_b^2 \xi)^{n+1}}\,.
\end{equation}
For the parametrisation of the spectral density we use the
reconstructed spectral density $R^{Rcstr}_b(s)$ which will be
discussed in the next section.
The continuum from open $B$ production sets in at 
$\sqrt{s}=2 M_{B}=10.56$ GeV just below the 4th resonance.
In the upsilon system the resonances are relatively dominant.
The start of the continuum threshold $s_0$ should thus in principle
be given by the mass of the 7th resonance. Nevertheless, when we take
into account only the first 3 resonances and a continuum threshold
$s_0$ typically 250 MeV above the 3th resonance and compare the result 
to an evaluation
with all 6 resonances and $s_0$ above the 6th resonance we miss
in the latter a contribution of 30\% from the continuum.
It seems natural to assume that this contribution originates
from open $B$ production. To account for this contribution we lower
the value of $s_0$ to $\sqrt{s_0}=11.0$ GeV. To estimate the error
we vary $s_0$ between $10.8\ \gev<\sqrt{s_0}<11.2\ \gev$.
To be conservative, in the analysis we also check the influence 
on the error if we remove the resonances above the continuum.

Above the charmonium threshold recent measurements have improved
the phenomenological situation significantly \cite{BES:01}.
85 data points have been taken in the region between 
$2.0\ \gev <\sqrt{s_0}<4.8\ \gev$ with an average precision of
6.6\%. The continuum threshold starts at $\sqrt{s}=2 M_{D}=3.73$ GeV.
From the measured spectral density the light quark contributions 
must be subtracted. At this energy the light
quarks can be safely assumed to be massless and the high energy
approximation \cite{c:97} provides a good description.
\begin{figure}
\begin{center}
\includegraphics[height=\figwidth,width=\figheight,angle=-90]{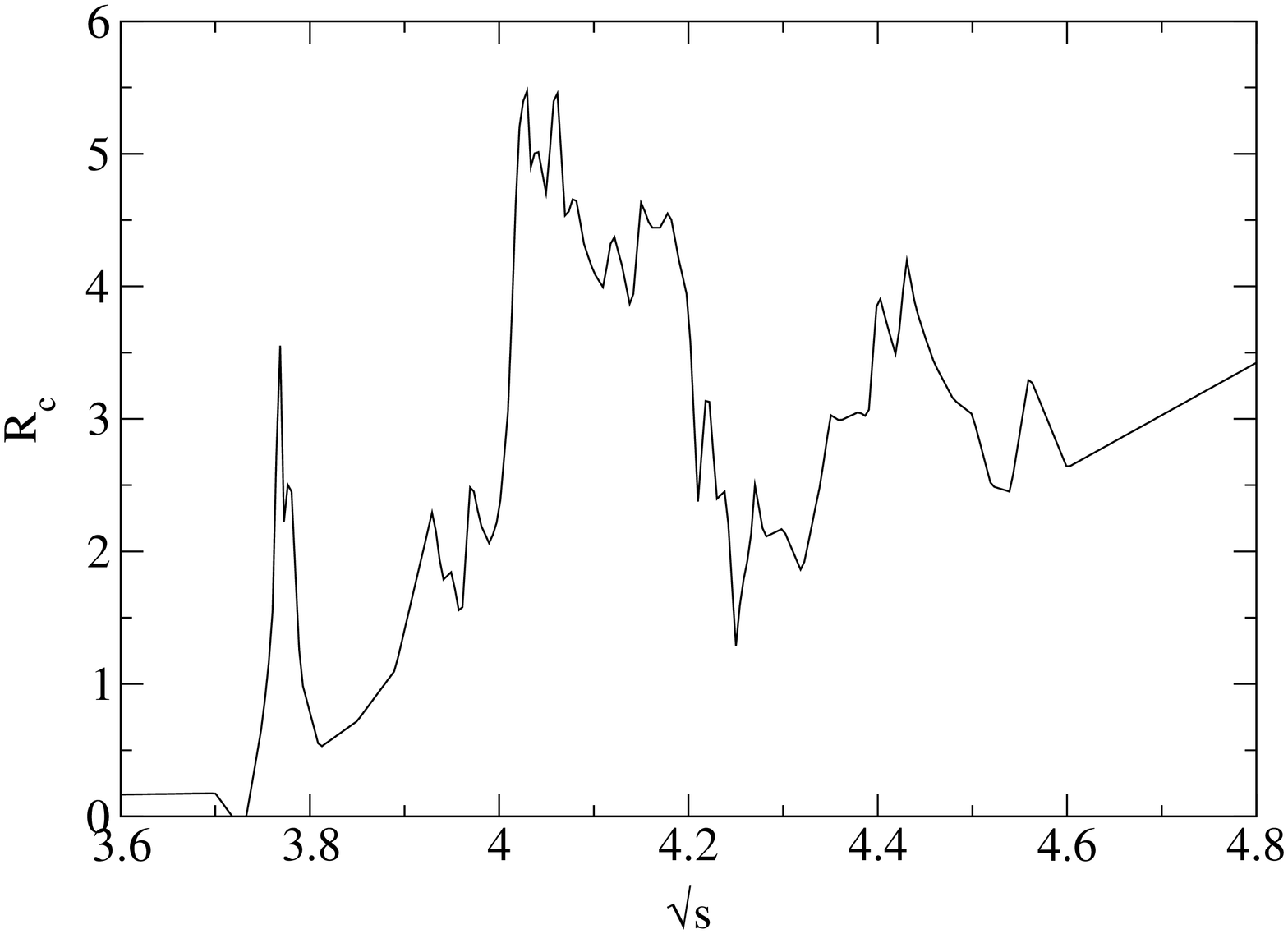}
\caption{\label{fig:6.a}$R_c(s)$
from \cite{BES:01}.}
\end{center}
\end{figure}
The resulting spectral density is shown in figure \ref{fig:6.a}.
At energies above the data points we again use the reconstructed spectral
density. Since the  resonances 3-6 are well reproduced by the data,
we only add the first two resonances below the continuum threshold.
The error from the data turns out
to be small compared to the theoretical uncertainties.

It is interesting to compare the measured cross section to the
predictions from quark-hadron-duality. In average, the reconstructed
spectral density lies above the data points. This should be no
surprise as the OPE demands an equality of the theoretical and
phenomenological moments only for the full correlator which also
includes the pole contributions. Since the lowest poles are
very dominant on the phenomenological side they are 
compensated by a larger theoretical spectral density for
intermediate values of $s$.
Consequently, one should take care when describing the 
phenomenological spectral density by the perturbative one, 
in particular, the choice
of the integration point $s_0$ could depend on the values
of $\xi$ and $n$. A more detailed description of the charmonium
cross section and the accuracy of quark-hadron-duality 
is presented in \cite{e:02}.
Now we explain how to construct a theoretical spectral density
for the full energy range.


\section{Reconstruction of the spectral density}

Besides the contributions from the poles of the Green's function 
and the condensates, the theoretical part of the correlator
contains the spectral density above threshold. Now we discuss
the different parts of the spectral density.

For high velocities the spectral density is well described
by the perturbative expansion.
As explained in section 4 we have used a method which allows a 
good approximation until relatively close to threshold.
The resummed spectral density, on the other hand, 
gives a good description for low values of $v$, but it fails to
describe the high energy part.
For these reasons, 
we introduce a separation velocity $v_{sep}$.
Above $v_{sep}$ we only use the perturbative spectral density.
Below $v_{sep}$ we essentially take the resummed spectral density.
The perturbative expansion has singular terms in $v$ which are
included in the resummed spectral density, but it also contains
contributions from higher powers in $v$ which can be isolated by
subtracting the double counted terms and these contributions
will be added to the resummed spectral density below $v_{sep}$.
\begin{figure}
\begin{center}
\includegraphics[height=\figwidth,width=\figheight,angle=-90]{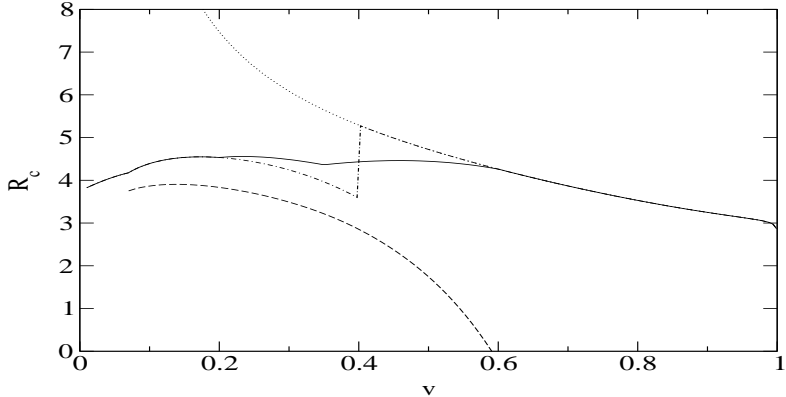}
\caption{\label{fig:7.a}$\psi$-system:
Solid line: interpolated spectral density;
dashed-dotted line: reconstructed spectral density;
dashed line: resummed spectral density;
dotted line: perturbative spectral density.}
\end{center}
\end{figure}
In fig. \ref{fig:7.a} we have displayed the charmonium spectral 
density from the different contributions as a function of $v$. This 
representation expands the threshold region. The dotted
line represents the perturbative expansion at NNLO. 
The dashed line is the resummed spectral density and 
the dashed-dotted line the reconstructed spectral density. 
For the charmonium system, there
exists a range of intermediate values of $v$ where neither the
perturbative expansion nor the resummation can be trusted. 
Indeed, it can be clearly seen that the reconstructed
spectral density shows a gap at the separation velocity.
Since this gap is not physical but a result of the mismatch between
the two energy regions we can try to construct a more physical
spectral density which interpolates smoothly between small
and large $v$. We can construct this interpolating spectral density from
$R^{Resum}$ and $R^{Pert}$ between the two velocities $v_1=0.2$ and 
$v_2=0.6$ with $R^{Inter}=R^{Resum}(v_2^2-v^2)/(v_2^2-v_1^2)
+R^{Pert}(v^2-v_1^2)/(v_2^2-v_1^2)$.
There is no explicit argument for a specific choice of $R^{Inter}$
except that it should give a smooth transition between the low and high energy
region. We have chosen a quadratic form instead of a linear one since
it suppresses better the behaviour of $R^{Resum}$ at high $v$
and of $R^{Pert}$ at low $v$. The moments from the interpolating spectral 
density are equal to the moments of the reconstructed spectral density
at $v_{sep}\approx 0.4$, a typical nonrelativistic velocity.
However, in the analysis we use the reconstructed spectral density
and vary $v_{sep}$ between 0.3 and 0.5 to estimate the error.
\begin{figure}
\begin{center}
\includegraphics[height=\figwidth,width=\figheight,angle=-90]{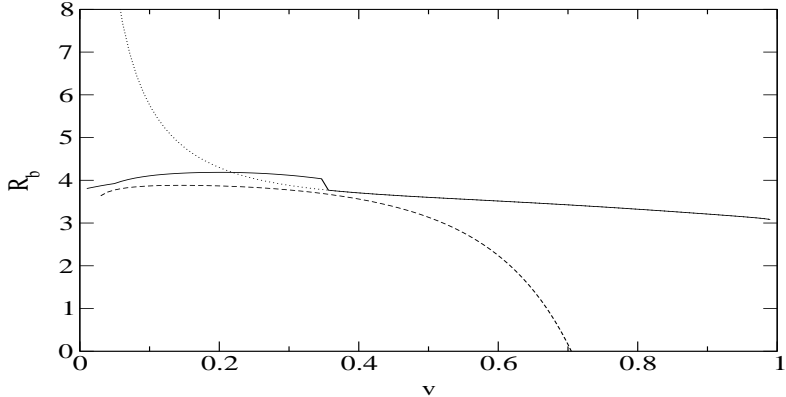}
\caption{\label{fig:7.b}$\Upsilon$-system:
Solid line: reconstructed spectral density;
dashed line: resummed spectral density;
dotted line: perturbative spectral density.}
\end{center}
\end{figure}

In figure \ref{fig:7.b} we have shown the same picture, but now
for typical scales of the upsilon system. Here the expansion behaves
much better. $R^{Resum}$ is valid up to higher and $R^{Pert}$ to lower 
velocities. So we obtain an overlap in the intermediate region
and the result is independent of variations around $v_{sep}\approx 0.35$.


\section{Numerical analysis for the bottom quark mass}

We now perform the analysis for the charm and bottom quark masses
in the pole and PS-scheme. Though the method of analysis will be 
similar in all four cases we discuss every case separately as each
requires a certain choice of parameters and an independent error analysis.
Since in the upsilon system the theoretical expansions converge better
we start with the bottom quark mass and devote the next section to the
charm quark mass.


\subsection{Pole mass scheme}

First, one has to fix the parameters on which the sum rule depends.
As a general rule, we will choose central values for the parameters
to determine the masses and then vary these parameters in appropriate
ranges for the error estimate. Let us start with the values
of $\xi$ and $n$. Since the bottom quark is relatively heavy, even for
$\xi=0$ the nonrelativistic and perturbative expansions converge
reasonably well. Nevertheless, the contributions from the poles of the
Green's function still dominate the theoretical part. To reduce their
influence and to spread the theoretical contributions more equally
among the poles, the resummed spectral density and the perturbative
spectral density we must choose a higher $\xi$. However, for 
$\xi>1$ the moments loose sensitivity on the mass and
the error from the input  parameters increases. Therefore we 
use a central value of
$\xi=0.5$ and vary $\xi$ between $0\leq \xi\leq 1$. 
Since the relevant scale for the evaluation point is the lowest
bound state energy, values of $\xi=0,0.5$ or 1 already correspond
to well separated evaluation points. In the section
on the PS-scheme we will justify this choice numerically as well.

As was shown in section 2, high values of $n$ enhance the threshold
region and the pole contributions. To keep the theoretical expansions
under control we restrict the moments to $n\leq 10$. From the
lower side, $n$ is limited by the phenomenological uncertainty.
Since for $n\le 4$ the continuum has a large influence on the mass
we use a range of $5\le n\le 10$.
As central values for our scales we have selected: 
\begin{equation}
  \label{eq:8.a}
  \mu_{soft}= 2.5 \ \mbox{GeV}\,, \quad \mu_{fac}= 3.5\ \mbox{GeV}\,,\quad
  \mu_{hard}= 5.0\ \mbox{GeV}\,. 
\end{equation}
We set the hard scale to the central value for the pole mass.
The soft scale should be given by the mass times a typical velocity.
We choose a value of $\mu_{soft}= 2.5$ GeV. Though 
one may prefer a slightly lower value of $\mu_{soft}$, 
the nonrelativistic expansion
gets large corrections for  $\mu_{soft}< 2.0$ GeV and we will use
this value as the lower bound in the variation of $\mu_{soft}$.
The factorisation scale separates the different regions and 
should lie between the two other scales. The selected scales are
required for a correct description of the spectral density.
Since this is a physical quantity the scales must be chosen independent of
$\xi$ and $n$ which merely serve as an evaluation point for the
moments.

As discussed in the section about the phenomenological spectral density,
we employ a continuum threshold of $\sqrt{s_0}=11.0\pm 0.2$ GeV. We use
a separation velocity of $v_{sep}=0.35$ and the result is independent
for a choice around this value. In the upsilon system the contribution
from the condensates is suppressed by roughly a factor of 
$10^{-3}$-$10^{-4}$
as compared to the pole contributions and can be safely neglected.
\begin{figure}
\begin{center}
\includegraphics[height=\figwidth,width=\figheight,angle=-90]{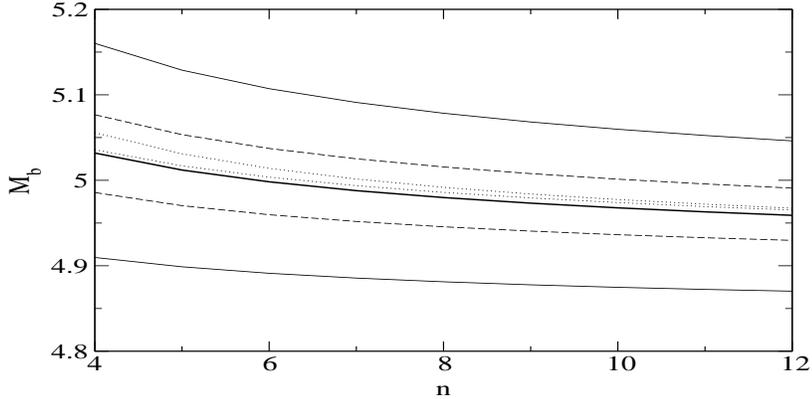}
\caption{\label{fig:8.a}
Thick solid line: central pole mass;
thin solid lines: $M_b$ for  $\mu_{soft}=2.0$
and 3.5 GeV;
dashed lines: $M_b$ for  $\mu_{fac}=2.0$
and 5.0 GeV;
dotted lines: $M_b$ for  $\mu_{hard}=2.5$
and 10.0 GeV.}
\end{center}
\end{figure}

\begin{table}
\begin{center}
\begin{tabular}{|c|c|c|c|c|c|c|}\hline
$n$  & 5 & 6 & 7 & 8 & 9 & 10 \\ \hline
${\cal M}_n^{Poles}$ & 0.079 & 0.055 & 0.038 & 0.026 & 0.018 & 0.013 \\ 
${\cal M}_n^{Rcstr}$ & 0.098 & 0.055 & 0.032 & 0.019 & 0.011 & 0.0067  \\ 
${\cal M}_n^{Continuum}$ & 0.047 & 0.023 & 0.012 & 0.0059 & 
0.0031 & 0.0016 \\ \hline
\end{tabular}
\caption{\label{tab:8.a}Moments for different $n$ with the parameters
$\xi=0.5$, $\mu_{soft} = 2.5$ GeV, $\mu_{fac}= 3.5$ GeV,
$\mu_{hard}= 5.0$ GeV,
$\sqrt{s_0}= 11.0$ GeV and $v_{sep}=0.35$.}
\end{center}
\end{table}

In table \ref{tab:8.a} we have collected the individual 
moments for different values of $n$. 
${\cal M}_n^{Poles}$ are the theoretical poles of the Green's function.
${\cal M}_n^{Rcstr}$ contains the moments from the reconstructed
spectral density and includes  the moments from the resummed spectral density
${\cal M}_n^{Resum}$ below and the perturbative spectral density
${\cal M}_n^{Pert}$ above $v_{sep}$. There is no clear distinction between 
${\cal M}_n^{Resum}$ and ${\cal M}_n^{Pert}$  since 
$v_{sep}$ can be used to shift their values.
${\cal M}_n^{Continuum}$ are the moments from the continuum part
of the reconstructed spectral density above $s_0$. For values
between $5\leq n \leq 10$ the moments ${\cal M}_n^{Rcstr}$ are of a
similar size as the pole contributions: for $n=5$ they exceed the poles
and become smaller for $n\geq 7$. 
The influence of the continuum
moments is relatively strong for small $n$ but gets more and more
suppressed for higher $n$. 
The thick solid line in figure \ref{fig:8.a}
shows the central value for the pole mass with the scales from
eq. \eqn{eq:8.a}. Averaging over the mass between $5\leq n\leq10$
we obtain
\begin{equation}
  \label{eq:8.a.2}
	M_b=4.984 \ \gev.
\end{equation}
The error originates mainly from the variation of the
scales. For the soft scale we choose a range of
$2.0 \ \gev \leq \mu_{soft}\leq 3.5\ \gev$. Below 
$\mu_{soft}< 2.0\ \gev$ the pole contributions show a bad perturbative
behaviour and the analysis becomes unstable. $\mu_{soft}> 3.5\ \gev$
would use a soft scale at an energy too high from physical expectation.
The hard scale is varied between $M_b/2$ and  $2 M_b$,
$2.5 \ \gev \leq \mu_{hard}\leq 10.0\ \gev$. As before for the central 
values, also the variation of the factorisation scale should lie 
between the two other scales and we use 
$2.0 \ \gev \leq \mu_{fac\ }\leq 5.0\ \gev$. In figure \ref{fig:8.a}
we have also plotted the change of the mass for a variation of these
scales. The error amounts to
\begin{eqnarray}
  \label{eq:8.b}
  2.0 \ \mbox{GeV} \leq \mu_{soft}\leq 3.5\ \mbox{GeV}:&&
  \Delta M_b=95\ \mbox{MeV}\,, \nn\\
  2.0 \ \mbox{GeV} \leq \mu_{fac\ }\leq 5.0\ \mbox{GeV}:&&
  \Delta M_b=35\ \mbox{MeV}\,, \nn\\
  2.5 \ \mbox{GeV} \leq \mu_{hard}\leq 10.0\ \mbox{GeV}:&&
  \Delta M_b=20\ \mbox{MeV}\,. 
\end{eqnarray}
\begin{table}
\begin{center}
\begin{tabular}{|c|c|c|c|c|c|c|}\hline
$\mu_{soft}$  & 2.0 & 2.25 & 2.5 & 2.75 & 3.0 & 3.5  \\ \hline 
${\cal M}_7^{Poles}$ & 0.050 & 0.043 & 0.038 & 0.034 & 0.032 & 0.028 \\
${\cal M}_7^{Rcstr}$ & 0.034 & 0.032 & 0.032 & 0.031 & 0.031 & 0.030 \\ \hline\hline
$\mu_{fac}$  & 2.0 & 3.0 & 3.5 & 4.0 & 4.5 & 5.0  \\ \hline
${\cal M}_7^{Poles}$ & 0.043 & 0.039 & 0.038 & 0.037 & 0.036 & 0.034  \\
${\cal M}_7^{Rcstr}$ & 0.032 & 0.032 & 0.032 & 0.032 & 0.031 & 0.031 \\ 
\hline \hline
$\mu_{hard}$  & 2.5 & 4.0 & 5.0 & 6.0 & 8.0 & 10.0  \\ \hline
${\cal M}_7^{Poles}$ & 0.038 & 0.038 & 0.038 & 0.038 & 0.039 & 0.039 \\ 
${\cal M}_7^{Rcstr}$ & 0.035 & 0.032 & 0.032 & 0.031 & 0.031 & 0.030 \\ 
\hline 
\end{tabular}
\caption{\label{tab:8.b}${\cal M}_7^{Poles}$ and ${\cal M}_7^{Rcstr}$
for different $\mu_{soft}$ with
$\mu_{fac}=3.5$ GeV and $\mu_{hard}=5.0$ GeV,
for different $\mu_{fac}$ with $\mu_{soft}=2.5$ GeV and
$\mu_{hard}=5.0$ GeV and
for different $\mu_{hard}$ with $\mu_{soft}=2.5$ GeV and
$\mu_{fac}=3.5$ GeV.} 
\end{center}
\end{table}
In table \ref{tab:8.b} we have listed the dependence of 
${\cal M}_7^{Poles}$ and ${\cal M}_7^{Rcstr}$
on the scales. The soft scale has a particular large influence
on the poles.
\begin{table}
\begin{center}
\begin{tabular}{|c|c|c|c|c|c|c|c|c|}\hline
$\mu_{soft}$ & & 2.0 & 2.25 & 2.5 & 2.75 & 3.0 & 3.5  \\ \hline 
 & LO & 0.024 & 0.020 & 0.018 & 0.015 & 0.014 & 0.012  \\ 
${\cal M}_7^{Poles}$ & NLO & 0.033 & 0.029 & 0.027 & 0.024 & 0.022 & 0.019  \\
 & NNLO & 0.050 & 0.043 & 0.038 & 0.034 & 0.032 & 0.028  \\ \hline
 & LO & 0.022 & 0.021 & 0.020 & 0.019 & 0.019 & 0.018  \\ 
${\cal M}_7^{Resum}$ & NLO & 0.014 & 0.014 & 0.014 & 0.014 & 0.014 & 0.014  \\
 & NNLO & 0.016 & 0.015 & 0.014 & 0.014 & 0.014 & 0.013  \\ \hline
\end{tabular}
\caption{\label{tab:8.c}Size of the moments
from the  poles and the resummed spectral density 
at LO, NLO and NNLO for different values of $\mu_{soft}$.}
\end{center}
\end{table}
In table \ref{tab:8.c} we have confronted the LO, NLO and NNLO corrections
in NRQCD for $n=7$. Instead of the ${\cal M}_7^{Rcstr}$
we directly use ${\cal M}_7^{Resum}$ for this comparison.
The analysis confirms that the expansions converge better for smaller
$n$ and behave worse for higher $n$ as is expected from the more 
sensitive testing of the threshold region.
The corresponding shift of the mass by going from NLO to NNLO amounts
to $\Delta M_b\approx 120\ \mev$.

Now we turn our attention to the other parameters. In the pole mass
scheme, a significant uncertainty comes from $\Lambda_{QCD}$ as well.
To determine $\alpha_s(\mu)$ we have used  $\alpha_s(M_Z)=0.1181
\pm 0.002$ \cite{pdg:00} and run this value down with the three-loop 
beta function. The corresponding $\Lambda_{QCD}$ for three loops
and four massless flavours is then $\Lambda_{QCD}=279\pm 29$ MeV.
Though $\alpha_s$ is known relatively precise, the error on the mass
amounts to $\Delta M_b=50$ MeV since the pole mass has a strong 
dependence on the coupling constant. The choice of the continuum 
threshold shifts the mass at low $n$ and gives an error of
$\Delta M_b=20$ MeV. The error from the experimentally measured
decay widths is $\Delta M_b=30$ MeV. When we employ a value of $\xi=0$
the mass decreases by 35 MeV and rises by the same amount for  $\xi=1$.
We have summarised the results in table \ref{tab:8.d}.
\begin{table}
\begin{center}
\begin{tabular}{|l|r|}\hline
\multicolumn{1}{|c}{Source} & \multicolumn{1}{|c|}{$\Delta M_b$} \\ \hline 
Variation of $\mu_{soft}$ & 95 MeV \\
Variation of $\mu_{fac}$ & 35 MeV \\
Variation of $\mu_{hard}$ & 20 MeV \\ 
Threshold $s_0$ & 20 MeV \\
Experimental widths & 30 MeV \\
Variation of $\Lambda_{QCD}$ & 50 MeV \\
Variation of $\xi$ & 35 MeV \\ \hline
Total error & 125 MeV \\ \hline 
\end{tabular}
\caption{\label{tab:8.d}Single contributions to the error of $M_b$.}
\end{center}
\end{table}
We have checked for the correlations between the errors from the different
input parameters and have
found almost no correlation between the errors. This holds also true for the
PS-scheme and for the charm quark mass analysis. Thus we add 
the errors quadratically and our final result for the pole mass is
\begin{equation}
  \label{eq:8.c}
  M_b=4.984 \pm 0.125 \ \mbox{GeV}\,.
\end{equation}
Using the  three-loop relation between the pole and the ${\rm \MSb}$-mass
which has been calculated recently \cite{mr:99,cs:99},
we obtain $m_{b}(m_{b})=4.277\pm 0.116$ GeV for the ${\rm \MSb}$-mass.
However, the relation between the two masses implicitly includes an
uncertainty of $O(\Lambda_{QCD})$.

Before turning to the analysis in the PS-scheme we want to investigate the
size of the NNNLO-corrections from NRQCD. In \cite{bpsv:99:2} the leading-log
term of $O(\alpha_s^3 \ln \alpha_s)$ to the energy levels has been 
derived in the framework
of potential NRQCD. The results have been confirmed in \cite{kp:00}
where also the corrections to the wave function at $O(\alpha_s^3 \ln^2 \alpha_s)$
were calculated. Recently, the full NNNLO corrections have been
computed \cite{kpss:02,ps:02}, but only for the energy levels itself.
The results of  \cite{kp:00} can be used to estimate the impact on the 
Green's function
and the mass. The main contributions come from the two lowest bound states.
With a typical soft scale of $\mu_{soft}=2.5$ GeV for the Green's function, 
the  $O(\alpha_s^3)$-contributions lower the mass by approximately 35 MeV, 
$\Delta M_b=-35\ \mev$. We have not included this mass shift in our final result 
since these contributions represent only a part of the full NNNLO corrections.
Furthermore, the results have only been derived for $|\psi_n(0)|^2$ and $E_n$
and not for the Green's function itself. Since the expansion for the energy and
the wave function is not very good, the mass shift may be overestimated. However,
it could indicate the size of the NNNLO corrections.


\subsection{Potential-subtracted mass scheme}

\begin{figure}
\begin{center}
\includegraphics[height=\figwidth,width=\figheight,angle=-90]{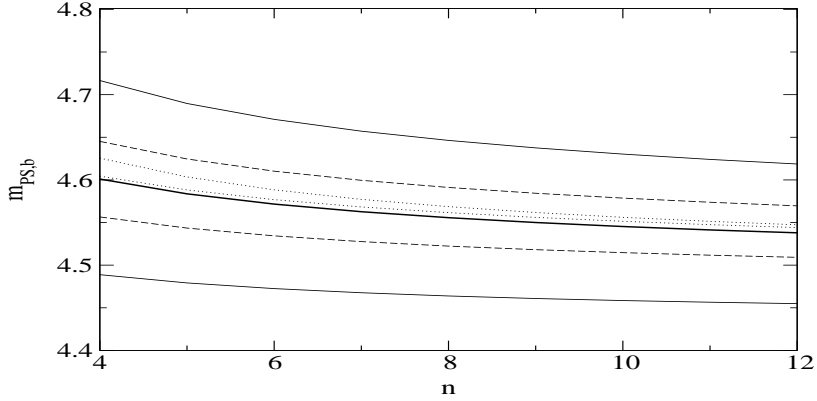}
\caption{\label{fig:8.b}
Thick solid line: central PS-mass;
thin solid lines: $m_{PS,b}$ for  $\mu_{soft}=2.0$
and 3.5 GeV;
dashed lines: $m_{PS,b}$ for  $\mu_{fac}=2.0$
and 5.0 GeV;
dotted lines: $m_{PS,b}$ for  $\mu_{hard}=2.5$
and 10.0 GeV.}
\end{center}
\end{figure}

Here the separation scale $\mu_{sep}$ appears as an additional parameter
which enters in the definition of the PS-mass \eqn{eq:3.g}.
This scale should be taken large enough in order to guarantee a perturbative
relation to the ${\rm \MSb}$-mass. On the other hand, it should be 
smaller than $Mv$ as not to affect the threshold behaviour. A good value
is $\mu_{sep}=2.0$ GeV and we will investigate a range of 
$1.0 \ \gev\leq\mu_{sep}\leq 3.0\ \gev$ to check the influence
on the ${\rm \MSb}$-mass. In figure \ref{fig:8.b} we have plotted the PS-mass
as a function of $n$. As our central values we obtain
\begin{equation}
  \label{eq:8.d}
	m_{PS,b}(\mu_{sep}=2\ \gev)=4.561\ \gev\,,\qquad
	m_b(m_b)=4.241 \ \gev\,.
\end{equation}
Relating the pole, PS- and ${\rm \MSb}$-masses
we make use of the recently calculated three-loop result for
the masses  \cite{mr:99,cs:99}. Since resummation includes all orders
in $\alpha_s$, this choice is more appropriate than the two-loop
relation. From the variations of the scales we obtain
\begin{eqnarray}
  \label{eq:8.d.2}
  2.0 \ \mbox{GeV} \leq \mu_{soft}\leq 3.5\ \mbox{GeV}:&& 
  \Delta m_{PS,b}=90\ \mbox{MeV}\,,\nn\\
  2.0 \ \mbox{GeV} \leq \mu_{fac\ }\leq 5.0\ \mbox{GeV}:&& 
  \Delta m_{PS,b}=35\ \mbox{MeV}\,,\nn\\
  2.5 \ \mbox{GeV} \leq \mu_{hard}\leq 10.0\ \mbox{GeV}:&& 
  \Delta m_{PS,b}=10\ \mbox{MeV}\,. 
\end{eqnarray}
\begin{table}
\begin{center}
\begin{tabular}{|c|c|c|c|c|c|c|}\hline
$n$  & 5 & 6 & 7 & 8 & 9 & 10 \\ \hline
${\cal M}_n^{Poles}$ & 0.043 & 0.027 & 0.017 & 0.010 & 0.0063 & 0.0039 \\ 
${\cal M}_n^{Rcstr}$ & 0.056 & 0.028 & 0.014 & 0.0076 & 0.0040 & 0.0022  \\ 
${\cal M}_n^{Continuum}$ & 0.025 & 0.011 & 0.0047 & 0.0021 & 
0.00095 & 0.00044 \\ \hline
\end{tabular}
\caption{\label{tab:8.e}Moments for different $n$ with the parameters
$\mu_{soft} = 2.5$ GeV, $\mu_{fac}= 3.5$ GeV,
$\mu_{hard}= 5.0$ GeV,
$\sqrt{s_0}= 11.0$ GeV and $v_{sep}=0.35$.}
\end{center}
\end{table}
In table \ref{tab:8.e} we again have collected the moments for
different values of $n$. For low $n$, ${\cal M}_n^{Rcstr}$ exceeds
${\cal M}_n^{Poles}$ whose influence grows for larger $n$.
\begin{table}
\begin{center}
\begin{tabular}{|c|c|c|c|c|c|c|}\hline
$\mu_{soft}$  & 2.0 & 2.25 & 2.5 & 2.75 & 3.0 & 3.5  \\ \hline 
${\cal M}_7^{Poles}$ & 0.022 & 0.019 & 0.017 & 0.015 & 0.014 & 0.012 \\
${\cal M}_7^{Rcstr}$ & 0.016 & 0.015 & 0.014 & 0.014 & 0.014 & 0.014 \\ \hline\hline
$\mu_{fac}$  & 2.0 & 3.0 & 3.5 & 4.0 & 4.5 & 5.0  \\ \hline
${\cal M}_7^{Poles}$ & 0.019 & 0.017 & 0.017 & 0.016 & 0.016 & 0.015  \\
${\cal M}_7^{Rcstr}$ & 0.015 & 0.015 & 0.014 & 0.014 & 0.014 & 0.014 \\ 
\hline \hline
$\mu_{hard}$  & 2.5 & 4.0 & 5.0 & 6.0 & 8.0 & 10.0  \\ \hline
${\cal M}_7^{Poles}$ & 0.017 & 0.017 & 0.017 & 0.017 & 0.017 & 0.017 \\ 
${\cal M}_7^{Rcstr}$ & 0.016 & 0.015 & 0.014 & 0.014 & 0.014 & 0.014 \\ 
\hline \hline
\end{tabular}
\caption{\label{tab:8.f}${\cal M}_7^{Poles}$ and ${\cal M}_7^{Rcstr}$
for different $\mu_{soft}$ with
$\mu_{fac}=3.5$ GeV and $\mu_{hard}=5.0$ GeV,
for different $\mu_{fac}$ with $\mu_{soft}=2.5$ GeV and
$\mu_{hard}=5.0$ GeV and
for different $\mu_{hard}$ with $\mu_{soft}=2.5$ GeV and
$\mu_{fac}=3.5$ GeV.} 
\end{center}
\end{table}
\begin{table}
\begin{center}
\begin{tabular}{|c|c|c|c|c|c|c|c|c|}\hline
$\mu_{soft}$ & & 2.0 & 2.25 & 2.5 & 2.75 & 3.0 & 3.5  \\ \hline 
 & LO & 0.011 & 0.0089 & 0.0077 & 0.0067 & 0.0061 & 0.0051  \\ 
${\cal M}_7^{Poles}$ & NLO & 0.015 & 0.013 & 0.012 & 0.011 & 0.0097 & 0.0084  \\
 & NNLO & 0.022 & 0.019 & 0.017 & 0.015 & 0.014 & 0.012  \\ \hline
 & LO & 0.010 & 0.0098 & 0.0094 & 0.0090 & 0.0087 & 0.0082  \\ 
${\cal M}_7^{Resum}$ & NLO & 0.0070 & 0.0070 & 0.0070 & 0.0070 & 0.0069 & 0.0068 \\
 & NNLO & 0.0077 & 0.0072 & 0.0068 & 0.0066 & 0.0064 & 0.0062  \\ \hline
\end{tabular}
\caption{\label{tab:8.g}Size of the moments
from the  poles and the resummed spectral density 
at LO, NLO and NNLO for different values of $\mu_{soft}$.}
\end{center}
\end{table}
In tables \ref{tab:8.f} and  \ref{tab:8.g} we have shown the scale
dependence and the behaviour for the different orders. 
The large coefficients which have been found in the Coulomb potential
at NNLO \eqn{ap:a.c} lead to a mass shift of $\Delta m_b(m_b)\approx
90-100\ \mev$ when going from NLO to NNLO.

When we lower the continuum threshold 
to $\sqrt{s_0}=10.8$ GeV, the mass decreases for small $n$.
For $\sqrt{s_0}=10.7$ GeV, the mass is completely stable for all values
of $3\leq n\leq 12$. But since the analysis is too complex to
draw conclusions on single parameters, we will not use the argument of 
stability to fix the values of the parameters. 
For higher values of $\sqrt{s_0}=11.2$ GeV the mass increases
for small $n$. The error from this variation is 
$\Delta m_{PS,b} =15$ MeV. Assuming that the continuum from open
B-production can be neglected, $\sqrt{s_0}$ should lie above the highest
resonance. Keeping the threshold at $\sqrt{s_0}=11.0$ GeV, we must
remove the 6th resonance. Then the mass increases by 6 MeV.
If we assume a very high threshold and also remove the 5th resonance
the mass rises by 22 MeV. The error from the experimental decay widths
is of a similar size, $\Delta m_{PS,b}
=25$ MeV. The main part comes from the width of the first resonance
though it is not as dominant as in the charmonium system.
In the PS-scheme, the influence of $\alpha_s$ is significantly
reduced. Using the same $\Lambda_{QCD}$ as before, we obtain 
$\Delta m_{PS,b}=20$ MeV for the PS-mass and $\Delta m_{b}(m_b)=5$ MeV
for the ${\rm \MSb}$-mass. From this weak dependence
on the strong coupling constant we can not use the analysis for an
estimate of $\alpha_s$.

Besides the contributions from resummation, this analysis includes the
perturbative spectral density above $v_{sep}$ as well. Therefore
it is interesting to investigate the influence of the perturbative part 
on the analysis. 
\begin{figure}[tbh]
\begin{center}
\includegraphics[height=\figwidth,width=\figheight,angle=-90]{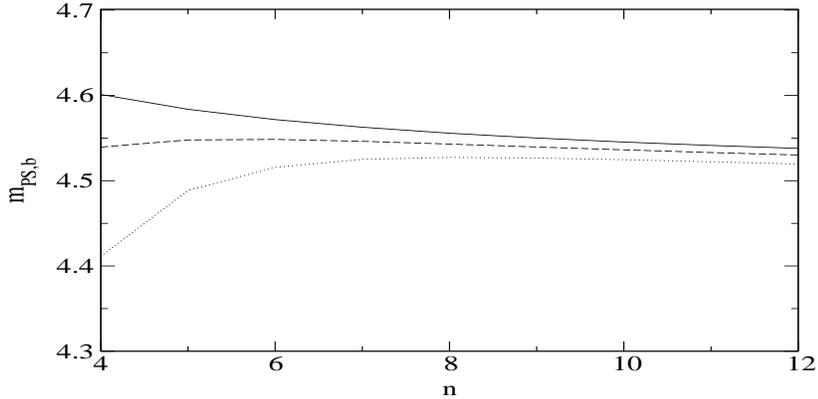}
\caption{\label{fig:8.c}
Solid line: central PS-mass;
dashed line: perturbative contribution only at LO;
dotted line: without perturbative contribution.}
\end{center}
\end{figure}
In figure \ref{fig:8.c} we have depicted the central mass
with a solid line. Then we keep all contributions, including the
poles, the resummed spectral density and the phenomenological part 
unchanged except for the perturbative
spectral density for which we use only the lowest order.
The resulting mass is shown as a dashed line. The analysis becomes
unstable for low $n$ since here the perturbative contributions play an
important part. Now we remove the perturbative part
completely and use only the resummed spectral density. The dotted
line signals clearly that essential information for low $n$ is lost.
Only for high $n$, dominated by the poles, the
analysis becomes more stable but here also the expansion of the
poles behaves more badly.

In table \ref{tab:8.h} we have varied the separation scale
$\mu_{sep}$ between $1.0\ \gev\leq \mu_{sep}\leq 3.0\ \gev$.
The stability of the PS-mass as a function
of $n$ remains almost unchanged, but the value of the PS-mass
changes strongly as its definition depends directly on $\mu_{sep}$.
In the relation to the ${\rm \MSb}$-mass this variation is cancelled
to such an extend that the ${\rm \MSb}$-mass changes only by 7 MeV.
\begin{table}
\begin{center}
\begin{tabular}{|c|c|c|c|c|c|}\hline
$\mu_{sep}$ & 1.0 & 1.5 & 2.0 & 2.5 & 3.0  \\ \hline 
$m_{PS,b}$ & 4.710 & 4.631 & 4.561 & 4.497 & 4.438 \\ 
$m_{b}(m_b)$ & 4.234 & 4.237 & 4.241 & 4.245 & 4.248 \\ \hline
\end{tabular}
\caption{\label{tab:8.h}Change of the masses for different
values of  $\mu_{sep}$.}
\end{center}
\end{table}

Now we want to discuss the choice of $\xi$ in more detail.
Using a higher $\xi$ the theoretical expansions converge better
and the dominance of the pole contributions is reduced.
As a result, the theoretical moments are more equally distributed.
Thus the dependence on a single contribution like the poles is reduced.
In this way one gets a better control over the systematic uncertainties
in the sum rules. 
\begin{table}
\begin{center}
\begin{tabular}{|c|c|c|c|}\hline
$\xi$ & $m_b(m_b)/[\gev]$ & $\Delta m_{b}^{(a)}/[\mev]$ &  
$\Delta m_{b}^{(b)}/[\mev]$ \\ \hline 
-0.25 & 4.200 & 71 & 114 \\ 
0 & 4.215 & 62 & 102 \\
0.5 & 4.241 & 54 & 90 \\
1.0 & 4.262 & 49 & 82 \\ 
1.5 & 4.278 & 46 & 79 \\
2.0 & 4.292 & 45 & 77 \\ \hline
\end{tabular}
\caption{\label{tab:8.i}Change of the ${\rm \MSb}$-mass, 
when adding the difference
from NNLO-NLO to the poles, $\Delta m_{b}^{(a)}$,
and the difference from NNLO-LO, $\Delta m_{b}^{(b)}$.}
\end{center}
\end{table}
Table \ref{tab:8.i} shows the ${\rm \MSb}$-mass for different $\xi$.
As a measure of the uncertainty
we now investigate the change of the mass connected with the expansion
of the poles in NRQCD. The first entry shows the central value for
the mass. Now we add to the poles of the Green's function the difference
between the NNLO and the NLO result. The increase of the mass is shown
in the second column. In the third column we add the difference between
the NNLO and the LO result.
Here the error on the mass from the expansion of
the poles decreases for higher $\xi$. If we assume this as a 
conservative error estimate, for $\xi=0.5$ this error is  of the same
order as the error from the variation of the soft scale.
In fact, the error from the scales increases for higher $\xi$ as
a result of the decreasing sensibility on the mass. From the viewpoint
of the convergence of the series, the scale variation tends to 
underestimate the error for low $\xi$ and to overestimate the error
for higher $\xi$. However, for $\xi=0.5$ both estimates are consistent
with each other and we have thus chosen this value as our default.
For $\xi\gsim 1$ the better control over the theoretical expansions
is not large enough to compensate for the decreasing sensibility on
the mass and the increasing influence
of the other contributions. The error from  $0\leq \xi\leq 1$
is $\Delta m_{PS,b}=30\ \mev$. 
\begin{table}
\begin{center}
\begin{tabular}{|l|r|r|} \hline
\multicolumn{1}{|c}{Source} & \multicolumn{1}{|c|}{$\Delta m_{PS,b}$} 
 & \multicolumn{1}{c|}{$\Delta m_b(m_b)$} 
\\ \hline 
Variation of $\mu_{soft}$ & 90 MeV & 80 MeV \\
Variation of $\mu_{fac}$ & 35 MeV  & 30 MeV\\
Variation of $\mu_{hard}$ & 10 MeV& 10 MeV \\ 
Variation of $\mu_{sep}$ & \multicolumn{1}{c|}{---}  & 5 MeV\\
Threshold $s_0$ & 25 MeV & 25 MeV\\
Experimental error & 35 MeV & 30 MeV\\
Variation of $\Lambda_{QCD}$ & 20 MeV& 5 MeV \\
Variation of $\xi$ & 30 MeV& 25 MeV \\ \hline
Total error & 112 MeV& 98 MeV \\ \hline 
\end{tabular}
\caption{\label{tab:8.j}Single contributions to the error 
of $m_{PS,b}$ and $m_b(m_b)$.}
\end{center}
\end{table}
Table \ref{tab:8.j} summarises
the error from all contributions for $m_{PS,b}$ and $m_b(m_b)$.
Our results are
\begin{eqnarray}
  \label{eq:8.e}
  m_{PS,b}(\mu_{sep}=2.0) &=& 4.561 \pm 0.112\ \mbox{GeV}\,,\nn\\
  m_{b}(m_{b}) &=& 4.241 \pm 0.098 \ \mbox{GeV}\,.
\end{eqnarray}
This value can be compared to the ${\rm \MSb}$-mass obtained from
the pole mass scheme, $m_{b}(m_{b})=4.277\pm 0.116\ \gev$.
The central value decreases by 30 MeV. In the PS-scheme one has 
better control over the systematic
uncertainties, reflected in an improved convergence for the 
theoretical expansions and in clear perturbative mass relation.


\section{Numerical analysis for the charm quark mass}


\subsection{Pole mass scheme}

The method of analysis will follow along the same lines as for the
bottom quark mass and we will put special  emphasis on the points different
in both analyses. First we must choose the value of $\xi$.
As for the bottomium, we will use $\xi=0.5$. At this value the pole
contributions still represent the dominant part. In principle one
would like to choose a higher value where the theoretical expansions
converge better. 
However, the contribution from the theoretical poles varies significantly
with the scales; for $\xi\gsim 1$ the mass depends too strongly on these
variations. Thus we again use a range of $0\leq \xi\leq 1$.
Again, in the PS-scheme we will investigate the perturbative
behaviour for different $\xi$ in more detail.
Since the perturbative expansions converge more slowly than for the upsilon
we restrict the analysis to smaller values of $n\leq 7$.
From the lower side, we choose $n\geq 4$ since for $\xi=0.5$ the moments at
$n=3$ already depend significantly on the phenomenological part. 
As central values for our scales we have selected: 
\begin{equation}
  \label{eq:9.a}
  \mu_{soft}= 1.2 \ \mbox{GeV}\,, \quad \mu_{fac}= 1.45\ \mbox{GeV}\,,\quad
  \mu_{hard}= 1.75\ \mbox{GeV}\,. 
\end{equation}
The hard scale corresponds to the central value of the pole mass.
For the soft scale we would have preferred a somewhat smaller value
but then the NNLO corrections become large.
\begin{table}
\begin{center}
\begin{tabular}{|c|c|c|c|c|c|c|}\hline
$n$  & 3 & 4 & 5 & 6 & 7 & 8 \\ \hline
${\cal M}_n^{Poles}$ & 0.65 & 0.48 & 0.35 & 0.26 & 0.19 & 0.14 \\ 
${\cal M}_n^{Rcstr}$ & 0.41 & 0.21 & 0.11 & 0.063 & 0.036 & 0.021  \\ 
${\cal M}_n^{Continuum}$ & 0.23 & 0.099 & 0.046 & 0.023 & 
0.011 & 0.0058 \\
${\cal M}_n^{Condensates}$ & -0.0033 & -0.0030 & -0.0027 & -0.0023 & 
-0.0019 & -0.0015 \\ \hline
\end{tabular}
\caption{\label{tab:9.a}Moments for different $n$ with the parameters
$\mu_{soft} = 1.2$ GeV, $\mu_{fac}= 1.45$ GeV,
$\mu_{hard}= 1.75$ GeV and $v_{sep}=0.4$.}
\end{center}
\end{table}
The moments for different values of $n$ are shown in table \ref{tab:9.a}.
${\cal M}_n^{Rcstr}$ are the moments from the reconstructed spectral
density at $v_{sep}=0.4$. At this separation velocity ${\cal M}_n^{Rcstr}$
equal the moments from the interpolating spectral density which was
introduced in section 7. The pole contributions dominate the
sum rule even for small $n$. The condensates are
suppressed compared to the poles and have no influence on the mass.
\begin{figure}
\begin{center}
\includegraphics[height=\figwidth,width=\figheight,angle=-90]{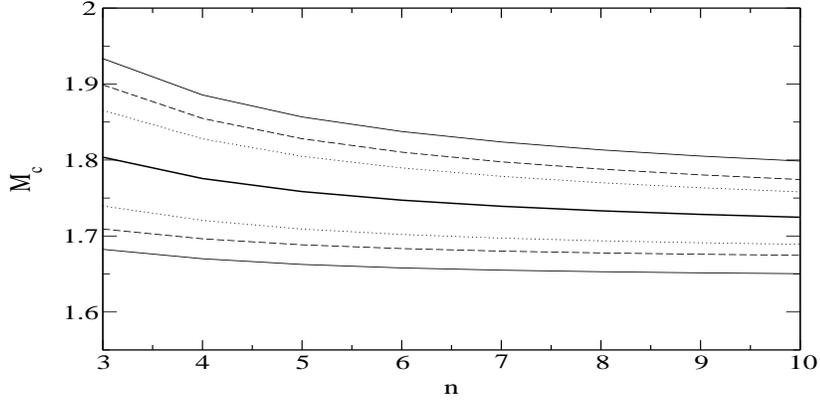}
\caption{\label{fig:9.a}
Thick solid line: central pole mass;
thin solid lines: $M_c$ for  $\mu_{soft}=1.1$
and 1.35 GeV;
dashed lines: $M_c$ for  $\mu_{fac}=1.2$
and 1.65 GeV;
dotted lines: $M_c$ for  $\mu_{hard}=1.4$
and 2.5 GeV.}
\end{center}
\end{figure}
From figure \ref{fig:9.a}, averaging over $4\leq n\leq 7$, we obtain
\begin{equation}
  \label{eq:9.a.2}
  M_c=1.754\ \gev\,.
\end{equation}
The error is dominated by the variation of the scales. 
For values of $\mu_{soft}\lsim 1.1\ \gev$ the pole contributions
get large NNLO corrections and we thus choose $1.1\ \gev\leq
\mu_{soft}\leq 1.35 \ \gev$. For the hard scale we use a range
of $1.4\ \gev\leq \mu_{hard}\leq 2.5 \ \gev$ and for the factorisation
scale  $1.2\ \gev\leq \mu_{fac}\leq 1.65 \ \gev$.
Since the convergence
of the nonrelativistic expansion is not very good for the 
charmonium system, the scales cannot be chosen arbitrarily 
far away from their central values.
Though the analysis is stable inside the given intervals,
the expressions tend to become unstable for scales outside of the
chosen ranges. The error amounts to
\begin{eqnarray}
  \label{eq:9.b}
  1.1 \ \mbox{GeV} \leq \mu_{soft}\leq 1.35\ \mbox{GeV}:&&
  \Delta M_c=90\ \mbox{MeV}\,, \nn\\
  1.2 \ \mbox{GeV} \leq \mu_{fac\ }\leq 1.65\ \mbox{GeV}:&&
  \Delta M_c=65\ \mbox{MeV}\,, \nn\\
  1.4 \ \mbox{GeV} \leq \mu_{hard}\leq 2.5\ \mbox{GeV}:&&
  \Delta M_c=40\ \mbox{MeV}\,. 
\end{eqnarray}
\begin{table}
\begin{center}
\begin{tabular}{|c|c|c|c|c|c|c|}\hline
$\mu_{soft}$  & 1.1 & 1.15 & 1.2 & 1.25 & 1.3 & 1.35  \\ \hline 
${\cal M}_5^{Poles}$ & 0.51 & 0.42 & 0.36 & 0.31 & 0.27 & 0.24 \\
${\cal M}_5^{Rcstr}$ & 0.12 & 0.11 & 0.11 & 0.11 & 0.10 & 0.10 \\ \hline\hline
$\mu_{fac}$  & 1.2 & 1.3 & 1.4 & 1.5 & 1.6 & 1.7  \\ \hline
${\cal M}_5^{Poles}$ & 0.46 & 0.42 & 0.38 & 0.34 & 0.29 & 0.25  \\
${\cal M}_5^{Rcstr}$ & 0.12 & 0.11 & 0.11 & 0.11 & 0.10 & 0.096 \\ 
\hline \hline
$\mu_{hard}$  & 1.4 & 1.6 & 1.8 & 2.0 & 2.2 & 2.5  \\ \hline
${\cal M}_5^{Poles}$ & 0.29 & 0.33 & 0.36 & 0.39 & 0.41 & 0.43 \\ 
${\cal M}_5^{Rcstr}$ & 0.11 & 0.11 & 0.11 & 0.11 & 0.11 & 0.11 \\ 
\hline 
\end{tabular}
\caption{\label{tab:9.b}${\cal M}_5^{Poles}$ and ${\cal M}_5^{Rcstr}$
for different $\mu_{soft}$ with
$\mu_{fac}=1.45$ GeV and $\mu_{hard}=1.75$ GeV,
for different $\mu_{fac}$ with $\mu_{soft}=1.2$ GeV and
$\mu_{hard}=1.75$ GeV and
for different $\mu_{hard}$ with $\mu_{soft}=1.2$ GeV and
$\mu_{fac}=1.45$ GeV.} 
\end{center}
\end{table}
\begin{table}
\begin{center}
\begin{tabular}{|c|c|c|c|c|c|c|c|c|}\hline
$\mu_{soft}$ & & 1.1 & 1.15 & 1.2 & 1.25 & 1.3 & 1.35  \\ \hline 
 & LO & 0.19 & 0.17 & 0.15 & 0.13 & 0.12 & 0.11  \\ 
${\cal M}_5^{Poles}$ & NLO & 0.31 & 0.28 & 0.25 & 0.23 & 0.21 & 0.19  \\
 & NNLO & 0.51 & 0.42 & 0.36 & 0.31 & 0.27 & 0.24  \\ \hline
 & LO & 0.097 & 0.093 & 0.090 & 0.087 & 0.085 & 0.083  \\ 
${\cal M}_5^{Resum}$ & NLO & 0.043 & 0.046 & 0.047 & 0.048 & 0.049 & 0.050  \\
 & NNLO & 0.045 & 0.040 & 0.036 & 0.033 & 0.031 & 0.029  \\ \hline
\end{tabular}
\caption{\label{tab:9.c}Size of the moments
from the  poles and the resummed spectral density 
at LO, NLO and NNLO for different values of $\mu_{soft}$.}
\end{center}
\end{table}
Tables \ref{tab:9.b} and \ref{tab:9.c} show the dependence of 
${\cal M}_5^{Poles}$ and ${\cal M}_5^{Rcstr}$ on the scales and of
${\cal M}_5^{Poles}$ and ${\cal M}_5^{Resum}$ from the different
orders on $\mu_{soft}$ respectively. 

To estimate the uncertainty
from $\alpha_s$ we employ $\Lambda_{QCD}=329\pm 29\ \mev$ which
is the corresponding value for 3 flavours and 3 loops. Then the mass
shifts by 60 MeV. Since already the lowest resonances dominate
the phenomenological part, the error from the measured spectral 
density and experimental widths is relatively small.
For $\xi=0$ the mass decreases by 60 MeV and increases by the same amount
for $\xi=1$. From table \ref{tab:9.d} we then obtain the pole mass
\begin{equation}
  \label{eq:9.c}
  M_c=1.754 \pm 0.147 \ \mbox{GeV}\,.
\end{equation}
\begin{table}
\begin{center}
\begin{tabular}{|l|r|}\hline
\multicolumn{1}{|c}{Source} & \multicolumn{1}{|c|}{$\Delta M_c$} \\ \hline 
Variation of $\mu_{soft}$ & 90 MeV \\
Variation of $\mu_{fac}$ & 65 MeV \\
Variation of $\mu_{hard}$ & 40 MeV \\ 
Experimental cross section & 5 MeV \\
Experimental widths & 20 MeV \\
Variation of $v_{sep}$ & 10 MeV \\
Variation of $\Lambda_{QCD}$ & 60 MeV \\
Variation of $\xi$ & 60 MeV \\ \hline
Total error & 147 MeV \\ \hline 
\end{tabular}
\caption{\label{tab:9.d}Single contributions to the error of $M_c$.}
\end{center}
\end{table}
This corresponds to a ${\rm \MSb}$-mass of $m_{c}(m_{c})=1.247\pm 0.134
\ \gev$. Again, there is an $O(\Lambda_{QCD})$ uncertainty
from the perturbative relation between the masses
so we now turn to the PS-scheme to determine the ${\rm \MSb}$-mass.


\subsection{Potential-subtracted mass scheme}

As in the pole scheme, we will use $\xi=0.5$ within a range of 
$0\leq \xi\leq 1$. For the separation scale we choose 
$\mu_{sep}=1.0\pm 0.2\ \gev$. This represents a compromise value.
It is still high enough for a perturbative evaluation and
sufficiently below the hard scale. 
Since in the PS-scheme the theoretical expansions converge better, 
one can employ a lower value for the soft scale
and we will use $\mu_{soft}=1.1$ GeV. As before, we
use a range of $4\leq n\leq 7$ for the moments.

In table \ref{tab:9.e} we have shown the moments for different $n$.
The poles represent the dominant part of the theoretical 
correlator. The size of the condensates is $\approx 1\%$ of the pole
contributions and also in this scheme they can be neglected for
the analysis.

Our central values for the PS- and  ${\rm \MSb}$-masses are
\begin{equation}
  \label{eq:9.d}
	m_{PS,c}(\mu_{sep}=1\ \gev)=1.300\ \gev\,,\qquad
	m_c(m_c)=1.188 \ \gev\,.
\end{equation}
\begin{figure}
\begin{center}
\includegraphics[height=\figwidth,width=\figheight,angle=-90]{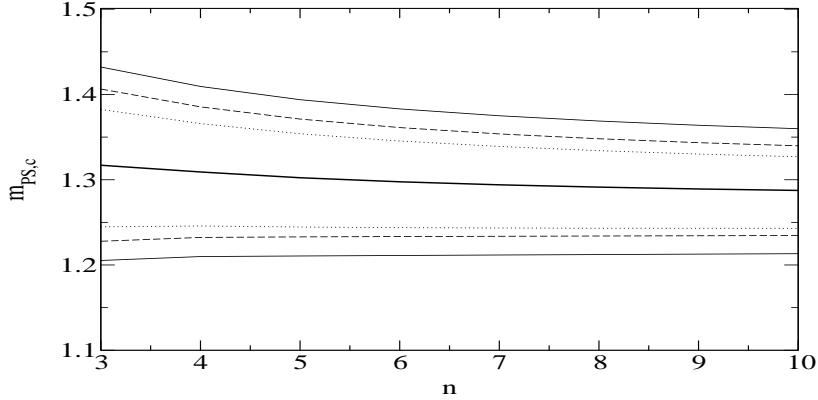}
\caption{\label{fig:9.b}
Thick solid line: central PS-mass;
thin solid lines: $m_{PS,c}$ for  $\mu_{soft}=1.0$
and 1.25 GeV;
dashed lines: $m_{PS,c}$ for  $\mu_{fac}=1.2$
and 1.65 GeV;
dotted lines: $m_{PS,c}$ for  $\mu_{hard}=1.4$
and 2.5 GeV.}
\end{center}
\end{figure}
\begin{table}
\begin{center}
\begin{tabular}{|c|c|c|c|c|c|c|}\hline
$n$  & 3 & 4 & 5 & 6 & 7 & 8 \\ \hline
${\cal M}_n^{Poles}$ & 0.22 & 0.11 & 0.056 & 0.028 & 0.014 & 0.0071 \\ 
${\cal M}_n^{Rcstr}$ & 0.13 & 0.043 & 0.016 & 0.0059 & 0.0022 & 0.00088  \\ 
${\cal M}_n^{Continuum}$ & 0.057 & 0.016 & 0.0047 & 0.0015 & 
0.00047 & 0.00015 \\
${\cal M}_n^{Condensates}$ & -0.0016 & -0.00097 & -0.00057 & -0.00032 & 
-0.00018 & -0.000096 \\ \hline
\end{tabular}
\caption{\label{tab:9.e}Moments for different $n$ with the parameters
$\mu_{soft} = 1.1$ GeV, $\mu_{fac}= 1.45$ GeV,
$\mu_{hard}= 1.75$ GeV and $v_{sep}=0.4$.}
\end{center}
\end{table}
In figure \ref{fig:9.b} we have plotted the PS-mass and the
corresponding error from the scales:
\begin{eqnarray}
  \label{eq:9.d.2}
  1.0 \ \mbox{GeV} \leq \mu_{soft}\leq 1.25\ \mbox{GeV}:&& 
  \Delta m_{PS,c}=85\ \mbox{MeV}\,,\nn\\
  1.2 \ \mbox{GeV} \leq \mu_{fac\ }\leq 1.65\ \mbox{GeV}:&& 
  \Delta m_{PS,c}=65\ \mbox{MeV}\,,\nn\\
  1.4 \ \mbox{GeV} \leq \mu_{hard}\leq 2.5\ \mbox{GeV}:&& 
  \Delta m_{PS,c}=50\ \mbox{MeV}\,.\nn\\ 
\end{eqnarray}
\begin{table}
\begin{center}
\begin{tabular}{|c|c|c|c|c|c|c|}\hline
$\mu_{soft}$  & 1.0 & 1.05 & 1.1 & 1.15 & 1.2 & 1.25  \\ \hline 
${\cal M}_5^{Poles}$ & 0.10 & 0.075 & 0.057 & 0.045 & 0.036 & 0.030 \\
${\cal M}_5^{Rcstr}$ & 0.020 & 0.017 & 0.015 & 0.013 & 0.012 & 0.011 \\ 
\hline\hline
$\mu_{fac}$  & 1.2 & 1.3 & 1.4 & 1.5 & 1.6 & 1.7  \\ \hline
${\cal M}_5^{Poles}$ & 0.083 & 0.072 & 0.062 & 0.052 & 0.043 & 0.034  \\
${\cal M}_5^{Rcstr}$ & 0.019 & 0.017 & 0.016 & 0.014 & 0.012 & 0.010 \\ 
\hline \hline
$\mu_{hard}$  & 1.4 & 1.6 & 1.8 & 2.0 & 2.2 & 2.5  \\ \hline
${\cal M}_5^{Poles}$ & 0.040 & 0.051 & 0.059 & 0.065 & 0.070 & 0.076 \\ 
${\cal M}_5^{Rcstr}$ & 0.013 & 0.014 & 0.015 & 0.016 & 0.017 & 0.017 \\ 
\hline 
\end{tabular}
\caption{\label{tab:9.f}${\cal M}_5^{Poles}$ and ${\cal M}_5^{Rcstr}$
for different $\mu_{soft}$ with
$\mu_{fac}=1.45$ GeV and $\mu_{hard}=1.75$ GeV,
for different $\mu_{fac}$ with $\mu_{soft}=1.1$ GeV and
$\mu_{hard}=1.75$ GeV and
for different $\mu_{hard}$ with $\mu_{soft}=1.1$ GeV and
$\mu_{fac}=1.45$ GeV.} 
\end{center}
\end{table}
\begin{table}
\begin{center}
\begin{tabular}{|c|c|c|c|c|c|c|c|c|}\hline
$\mu_{soft}$ & & 1.0 & 1.05 & 1.1 & 1.15 & 1.2 & 1.25  \\ \hline 
 & LO & 0.035 & 0.027 & 0.022 & 0.018 & 0.015 & 0.013  \\ 
${\cal M}_5^{Poles}$ & NLO & 0.064 & 0.050 & 0.040 & 0.033 & 0.028 & 0.024  \\
 & NNLO & 0.10 & 0.075 & 0.057 & 0.045 & 0.036 & 0.030  \\ \hline
 & LO & 0.019 & 0.016 & 0.014 & 0.013 & 0.011 & 0.010  \\ 
${\cal M}_5^{Resum}$ & NLO & 0.011 & 0.010 & 0.0097 & 0.0092 & 
0.0086 & 0.0081  \\
 & NNLO & 0.0096 & 0.0075 & 0.0061 & 0.0051 & 0.0043 & 0.0037  \\ \hline
\end{tabular}
\caption{\label{tab:9.g}Size of the moments
from the  poles and the resummed spectral density 
at LO, NLO and NNLO for different values of $\mu_{soft}$.}
\end{center}
\end{table}
Table \ref{tab:9.f} and \ref{tab:9.g} show the scale dependence and
the behaviour for the different orders.  
Compared to the pole scheme, the uncertainty on $\alpha_s$ is much
improved and amounts to $\Delta m_{PS,c}=20\ \mev$ and 
$\Delta m_c(m_c)=10\ \mev$.
\begin{figure}[tbh]
\begin{center}
\includegraphics[height=\figwidth,width=\figheight,angle=-90]{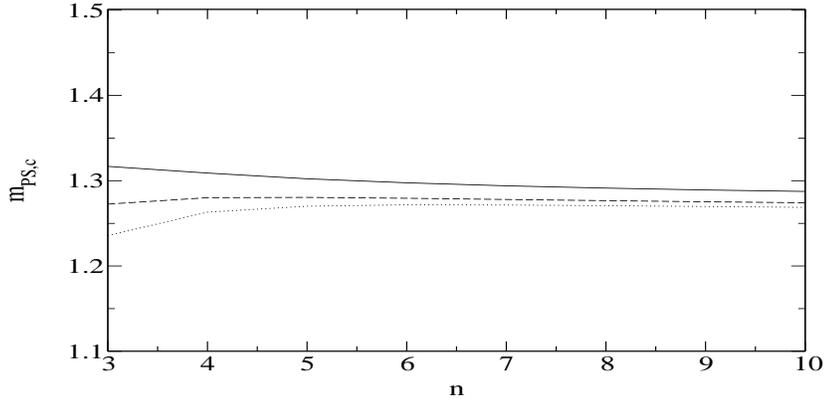}
\caption{\label{fig:9.c}
Solid line: central PS-mass;
dashed line: perturbative contribution only at LO;
dotted line: without perturbative contribution.}
\end{center}
\end{figure}
As for the bottomium we now investigate the 
significance of the perturbative contribution. The solid line
in figure \ref{fig:9.c} shows the central mass. Then we just
change the perturbative spectral density. The dashed line shows
the LO result and in the dotted line we neglect the perturbative
contribution completely. For low $n$ we loose stability for the mass
though the effect is not as pronounced as in the bottom case since
for the charmonium the poles play a more dominant part.

\begin{table}
\begin{center}
\begin{tabular}{|c|c|c|c|c|c|}\hline
$\mu_{sep}$ & 0.8 & 0.9 & 1.0 & 1.1 & 1.2  \\ \hline 
$m_{PS,c}$ & 1.353 & 1.326 & 1.300 & 1.277 & 1.255 \\ 
$m_{c}(m_c)$ & 1.181 & 1.184 & 1.188 & 1.192 & 1.196 \\ \hline
\end{tabular}
\caption{\label{tab:9.h}Change of the masses for different
values of  $\mu_{sep}$.}
\end{center}
\end{table}
The influence of the separation scale on the masses is shown in
table \ref{tab:9.h}. The definition of the PS-mass depends
directly on $\mu_{sep}$, but the ${\rm\MSb}$-mass remains very stable
and changes only by $\Delta m_c(m_c)=8\ \mev$. 

\begin{table}
\begin{center}
\begin{tabular}{|c|c|c|c|}\hline
$\xi$ & $m_c(m_c)/[\gev]$ & $\Delta m_{c}^{(a)}/[\mev]$ &  
$\Delta m_{c}^{(b)}/[\mev]$ \\ \hline 
-0.5 & 1.161 & 44 & 73 \\ 
0 & 1.174 & 35 & 64 \\
0.5 & 1.188 & 31 & 59 \\
1.0 & 1.205 & 30 & 56 \\ 
1.5 & 1.223 & 29 & 54 \\
2.0 & 1.241 & 30 & 54 \\ 
3.0 & 1.273 & 31 & 54 \\ 
4.0 & 1.300 & 34 & 56 \\ \hline
\end{tabular}
\caption{\label{tab:9.i}Change of the ${\rm \MSb}$-mass, 
when adding the difference
from NNLO-NLO to the poles, $\Delta m_{c}^{(a)}$,
and the difference from NNLO-LO, $\Delta m_{c}^{(b)}$.}
\end{center}
\end{table}
Now we turn our attention to the choice of $\xi$. In table \ref{tab:9.i}
the  ${\rm \MSb}$-mass is depicted for different $\xi$. As we have
done in table \ref{tab:8.i} we add to the pole contributions the
difference from the NNLO-NLO and from NNLO-LO. The change in the 
${\rm \MSb}$-mass is shown in the third and fourth column.
Since the poles are relatively dominant, in principle we would like
to use a higher value of $\xi$. But from table \ref{tab:9.i}
one can see that the better expansion is almost compensated
by the decreasing sensitivity for $\xi\gsim 1$. In addition,
the errors from the other input parameters grow. Like in the
bottom case, the error from the scales improves for lower $\xi$.
The error from the scales is still larger than the estimated uncertainty
from the expansion, but in order to be conservative we will use the
larger error from the scales for our error estimate.
The variation of $\xi$ changes the central value for the mass as well
and for $0\leq \xi\leq 1$ we obtain
$\Delta m_{PS,c}=20\ \mev$ which shows a much better 
behaviour than in the pole scheme.
\begin{table}
\begin{center}
\begin{tabular}{|l|r|r|} \hline
\multicolumn{1}{|c}{Source} & \multicolumn{1}{|c|}{$\Delta m_{PS,c}$} 
 & \multicolumn{1}{c|}{$\Delta m_c(m_c)$} 
\\ \hline 
Variation of $\mu_{soft}$ & 85 MeV & 75 MeV \\
Variation of $\mu_{fac}$ & 65 MeV  & 55 MeV\\
Variation of $\mu_{hard}$ & 50 MeV& 40 MeV \\ 
Variation of $\mu_{sep}$ & \multicolumn{1}{c|}{---}  & 10 MeV\\
Experimental cross section & 5 MeV & 5 MeV\\
Experimental widths & 20 MeV & 20 MeV\\
Variation of $v_{sep}$ & 10 MeV& 10 MeV \\
Variation of $\Lambda_{QCD}$ & 20 MeV& 10 MeV \\
Variation of $\xi$ & 20 MeV& 15 MeV \\ \hline
Total error & 124 MeV& 106 MeV \\ \hline 
\end{tabular}
\caption{\label{tab:9.j}Single contributions to the error 
of $m_{PS,c}$ and $m_c(m_c)$.}
\end{center}
\end{table}
A summary of all contributions to the error is presented in table \ref{tab:9.j}.
Finally we obtain the masses:
\begin{eqnarray}
  \label{eq:9.e}
  m_{PS,c}(\mu_{sep}=1.0) &=& 1.300 \pm 0.124\ \mbox{GeV}\,,\nn\\
  m_{c}(m_{c}) &=& 1.188 \pm 0.106 \ \mbox{GeV}\,.
\end{eqnarray}
This value is 59 MeV lower than the central value from the pole scheme.
This is no surprise since in the pole scheme the theoretical contributions
have large perturbative corrections and the relation between the masses
contains large uncertainties as well.

In our previous work on the charm quark mass \cite{ej:01} we obtained a slightly
higher charm quark mass of $m_{c}(m_{c}) = 1.23 \pm 0.09 \ \mbox{GeV}$. In this work
we have chosen an evaluation point of $\xi=0.5$. In the
theoretical QCD calculation of the moments in the PS-scheme we have now set the
start of the continuum in eq. \eqn{eq:3.f} according to the pole mass
which is the appropriate mass definition for free quark production. This leads
to a reduction of $m_c$ by 25 MeV. On the phenomenological side we have
included the BES data. This gives a better control in the region between 3.8
and 4.6 GeV where the assumptions of quark-hadron-duality cannot be expected to
work well. Finally we have extended the error analysis.


\section{Comparison to other mass determinations}

Now we compare our value for the charm quark mass to other determinations.
In this and the next section the ${\rm \MSb}$-masses are
always evaluated at their own scale, $m_c=m_c(m_c)$ and
$m_b=m_b(m_b)$.
The basis of the charmonium sum rules was already laid in
\cite{nosvvz:77,nosvvz:78}. Since then, many researchers have
extracted the charm quark mass from the sum rules. In \cite{dgp:94} 
the pole mass was determined from perturbation theory to NLO
resulting in a value of $M_c=1.46\pm 0.07$ GeV. In a second investigation
\cite{n:94:2,n:98} the analysis has been performed
in the ${\rm \MSb}$-scheme with perturbation theory to NLO. 
Using the NLO relation to the pole mass the author obtains
$m_c= 1.26\pm 0.05$ GeV and $M_c = 1.42 \pm 0.03$ GeV.
The author has also performed an analysis using resummation
in LO with a value of $M_c = 1.45 \pm 0.07$ GeV.
In our analysis the increased value of the pole mass is essentially
due to large Coulomb contributions which have not been included in
former analyses. As a consequence, the error becomes larger as well.

The charm quark mass can also be derived from direct application
of NRQCD to hadronic bound states. The authors of \cite{bsv:01,bsv:02}
have studied the energy level of the charmonium ground state.
They conclude a ${\rm \MSb}$-mass of $m_c=1.241\pm 0.015$ GeV
where the error is from the variation of $\alpha_s$ only. 
In \cite{py:98} a similar analysis was performed for the pole
mass with the result $M_c = 1.88 ^{+0.22}_{-0.13}$ GeV. 
In \cite{p:01} NRQCD was applied to the mass difference
between the $B$ and $D$ mesons with $m_c=1.21\pm 0.11$ GeV.
Further improvement of these determinations may be possible in the 
near future. However, these determinations face the problem that the
contributions from NRQCD must be directly evaluated at low
energies close to threshold
whereas in the sum rules the theoretical expansions can be evaluated
in a perturbative region.
Furthermore, nonperturbative effects
may have a significant impact on the charmonium energy levels.

During the last years several lattice analyses have been performed
for the charm quark mass \cite{accflm:94,k:98,ggrt:99,bf:96,bf:97,b:02}
with rather widespread results.
The most recent one \cite{b:02} obtains $m_c=1.26\pm0.04\pm 0.12$ GeV. 
This calculation was done at a lattice spacing of $a\approx 0.07$ fm.
Though it was done in quenched QCD, the authors expect a minor
decrease of 5\% for the unquenched case. Here also a discussion
on the previous lattice analyses can be found.
Very recently, two preliminary results \cite{rs:02,j:02} from different
lattice groups have been presented with 
$m_c=1.314\pm0.05$ GeV
and  $m_c=1.27\pm0.05$ GeV (statistical error only) respectively.	

Ref. \cite{n:01} has applied
pseudoscalar sum rules to heavy-light quark systems. Values of
$m_c=1.10\pm0.04$ GeV and $m_b=4.05\pm0.06$ GeV were deduced.
A discussion on this work can be found in \cite{jl:02}.

In some recent works the charm quark mass has been determined from
charmonium sum rules on perturbative grounds without 
Coulomb resummation.
The authors of \cite{ks:01} have used moment sum rules for the 
charmonium. They use moments of $n\approx 1-4$ and argue that
resummation is not necessary for such low values of $n$.
On the theoretical side $\Pi(s)$ is 
calculated up to $O(\alpha_s^2)$ at a scale of $\mu=3\ \gev$.
This is compared to the phenomenological part which includes
the $J/\psi$ and $\psi'$ resonances and the data from 
BES \cite{BES:01} above the $D$ threshold. They obtain
$m_c=1.30\pm0.03$ GeV. A similar analysis for the bottom yields
$m_b=4.21\pm0.05$ GeV.
In \cite{ps:01} a contour integration was performed to apply
the Cauchy sum rules. The integral was closed at an energy of 5 GeV.
At this scale the high energy approximation was used to calculate
$\Pi(s)$ up to $O(\alpha_s^2)$. As in \cite{ks:01}, this was
compared to the two lowest $\psi$-resonances plus the continuum
data \cite{BES:01}. The result is $m_c=1.37\pm 0.09$ GeV.
Very recently, the authors of \cite{el:02} obtain 
$m_c=1.289^{+0.040}_{-0.045}$ GeV from a comparison of the perturbative
spectral density to continuum data.
Ref. \cite{iz:02} has presented an update of the SVZ sum rules
\cite{svz:79} with perturbation theory at NNLO. They
extract the gluon condensate and a charm quark mass of
$m_c=1.275\pm0.015$ GeV.

However, in light of the present work it seems doubtful that
a reliable determination of the charm quark mass from the charmonium system
can be achieved by a pure perturbative evaluation without resummation
as in \cite{ks:01,ps:01,el:02,iz:02}. 
To clarify this point, let us return to the 
numerical analysis.
Choosing $\xi$ sufficiently large and $n$ small enough, one can
easily approach a region where the perturbative contribution represents 
the dominant part. One could then expect that the use of
perturbation theory and its relative small scale dependence
will give a reasonable approximation. Nevertheless, this conclusion is 
misleading.

For values of $\xi=4$ and $n=4$ the moment from the perturbative
contribution exceeds the poles by 30\% (for $n\le 3$ the phenomenological
continuum is very dominant and the analysis becomes unstable). Nevertheless,
a variation of the soft scale gives an error of $\Delta m_c=110\ \mev$
whereas the error from the hard scale is $\Delta m_c=70\ \mev$ and
this error could even be improved by using a high energy approximation.
For even higher $\xi$ the result is similar: for $\xi=6$ and $n=4$
the perturbative contribution is twice as large as the pole contributions,
but the error from $\mu_{soft}$ gives  $\Delta m_c=125\ \mev$ and
from the error from $\mu_{hard}$ amounts to $\Delta m_c=90\ \mev$. 
Though the pole contributions
are relatively suppressed, the mass reacts much stronger on the
remaining uncertainties. This behaviour can already be seen in
table \ref{tab:9.i}. Ever higher $\xi$ do not improve the accuracy
any more but in same way as the theoretical expansion improves the
sensitivity on the mass is lost. 
Indeed, if we set the pole contribution to zero in our analysis,
$m_c$ would drop by approximately 300 MeV even for very high $\xi$
and the analysis could not be trusted any more.

That a description without inclusion of the
theoretical poles is insufficient can already be seen from the
quantum mechanical sum rules for the mass \eqn{eq:2.v}.
On the phenomenological side, the
main dependence on the mass originates already from the first
bound states, even if the continuum part dominates the moments.
The contributions from the poles, starting from
$O(\alpha_s^3)$, must be included in the theoretical description
as well to obtain a reliable mass determination.
As was discussed in \cite{iz:02}, the charmonium system is not well
described as a Coulomb system. In particular, the expansion for the
higher states cannot be trusted and the effective potential may
differ from the Coulombic one. However, it is indispensable to use
the terms from resummation for a determination of the mass.
The most important contribution to the sum rules originates already
from the ground state (\ref{eq:2.u},\ref{eq:2.v}). The quantum mechanical sum rules
show clearly that fixed-order perturbation theory in a system whose
ground state is governed by a Coulomb-similar potential leads to an
instable and unreliable sum rule for the mass.

However, it should be kept in mind that a Coulomb-dominated description of the
charmonium system stands on less firm grounds than for the upsilon system whose
energy is sufficiently large to allow for a reliable resummation. This is also
reflected in the fact that the relative error of $m_c$ is almost a factor 4
larger than the relative error of $m_b$.

During the last years much effort has been dedicated to the determination 
of the bottom quark mass. The methods which have been employed were mainly 
based on QCD sum rules for the upsilon system, NRQCD for the bound states
or lattice QCD. We have listed some of these results in table
\ref{tab:10.a}.
\begin{table}
\begin{center}
\begin{tabular}{|l|l|l|} \hline
\multicolumn{3}{|c|}{QCD sum rules}\\ \hline 
Authors & $M_b/[GeV]$ & $m_b/[GeV]$ \\ \hline
V  \cite{v:95} & $4.83\pm 0.01$ & \\
KPP \cite{kpp:98} & $4.75\pm 0.04$ & \\
MY \cite{my:98} & & $4.20\pm 0.10 $\\
PP \cite{pp:99} & $4.80\pm 0.06$ & $4.21\pm 0.11$ \\
JP \cite{jp:97,jp:98} & $4.84\pm 0.08$ & $4.19\pm 0.06$  \\
BS \cite{b:99:2} & & $4.25\pm 0.08$  \\
H \cite{h:99,h:00,h:01} & & $4.17\pm 0.05$\\ 
This work & $4.98\pm 0.125$ & $4.24 \pm 0.10$ \\ \hline
\multicolumn{3}{|c|}{NRQCD potential}\\ \hline 
PY \cite{py:98} & $5.015^{+0.11}_{-0.07}$ & $4.45^{+0.05}_{-0.03}$ \\
BSV \cite{bsv:01,bsv:02} & & $4.19\pm 0.03$\\
P \cite{p:01} & & $4.21 \pm 0.09$ \\ \hline
\multicolumn{3}{|c|}{Lattice QCD}\\ \hline 
A et al. \cite{aetal:00} & & $4.35\pm 0.23$ \\ 
GGMR \cite{ggmr:00} & & $4.26\pm 0.09$ \\ \hline
\end{tabular}
\caption{\label{tab:10.a}Some references to the bottom quark mass.}
\end{center}
\end{table}
A more complete list of references can be found in \cite{pdg:00}.
We can directly compare our work to previous sum rule analyses.
As detailed discussions about the advances and drawbacks of
these analyses can be found in \cite{h:99,b:99:2,h:02},
here we rather want to point out some interesting differences.
For the comparison we will use the work by Hoang \cite{h:99,h:00,h:01} where
the most extensive analysis has been presented. 
In that moment sum rules the theoretical moments were directly expanded
for small energies around threshold. With help of a contour integration
this could be used to calculate the moments via the inverse Laplace
transform. Since the approach has focused on the nonrelativistic
properties of the upsilon system, for a comparison we now set the 
perturbative contributions in our analysis to zero. 
For a closer comparison we choose a value of $\xi=0$ where
the analysis of Hoang was performed. Two effects become important:
The central value of the mass decreases by 25 MeV. Second, at $\xi=0$
the poles are more dominant and the influence of the perturbative
contributions is reduced. For $\xi=0$, without perturbation theory
and for $7\le n\le 10$ we obtain $m_b=4.17\ \gev$. For $n\lsim 6$ the
analysis becomes unstable. Three differences remain.
For our central mass we have used a factorisation scale of 
$\mu_{fac}=3.5\ \gev$ whereas Hoang has performed a scan over 
$2.5\ \gev \le \mu_{fac}\le 10.0\ \gev$ which roughly corresponds
to a central value of $\mu_{fac}=5.0\ \gev$. Second, for the parametrisation
of the phenomenological continuum Hoang has used a continuum threshold
of $\sqrt{s_0}=10.56\ \gev$ which corresponds to the start of
$B\bar{B}$ production. As discussed in chapter 6, we have chosen
a value of  $\sqrt{s_0}=11.0\ \gev$ to parametrise the non-resonant
part of the spectral density. Using $\mu_{fac}=5.0\ \gev$ and 
$\sqrt{s_0}=10.56\ \gev$ we finally arrive at $m_b=4.14\ \gev$.
In his work, Hoang has estimated the effect of a finite charm quark mass
to be $\Delta m_b\approx -30\ \mev$. For a massless charm quark
he obtains $m_b=4.20\ \gev$ which is 60 MeV higher than our result
for similar input parameters.
Our error is larger than his one by a factor of two. Hoang has used
a $\chi^2$-fit with several moments. In this way he gets a cancellation
between theoretical contributions of the $\chi^2$-function. In our
analysis we keep the mass as a function of $n$
which serves as additional check for the stability of the sum rules.
Furthermore, at $\xi=0.5$ the mass reacts stronger
on a variation of the parameters. But as was discussed in the 
numerical analysis, in this way the control over the analysis
is improved and we believe that we have thus reduced the 
systematic uncertainties of the method.

Now we want to comment on a recent work on massless contributions
to the heavy quark correlator \cite{gp:01,gp:01:2}. 
Here it was shown that at $O(\alpha_s^3)$ the correlator contains
a three-gluon massless intermediate state. Its contribution
to the moments ${\cal M}_n(\xi=0)$ contains a divergent term
for $s=0$ and $n\ge 4$. Thus the authors have concluded that the 
moment sum rules
can only be reliably evaluated at $n<4$. We believe that this claim
is unfounded as was already noticed in \cite{nosvvz:78}. 
First, up to now the sum rule analyses contain 
perturbative contributions
up to $O(\alpha_s^2)$ and higher order terms from resummation. At this
order the 3-gluon cut was not and should not be included in the analysis
and all quantities are well defined. Furthermore, even if the calculation was
including the full $O(\alpha_s^3)$ contributions, these terms must not
be taken into account. The terms from the three gluon cut mainly
correspond to light quark production and the divergent parts at
$s=0$ are entirely due to light quark production as the heavy quark pair
only gives a contribution above $s>4 M^2$.
Since they are not included
in the phenomenological part they also should not appear in the 
theoretical part and must be explicitly subtracted from the perturbative
contributions. The remaining ambiguity which results from the difficulty
to separate the light and heavy quark production in the dispersion
integral above $s>4 M^2$ is a finite effect of  $O(\alpha_s^3)$
and can be completely neglected within the uncertainty of this work.

As discussed in \cite{pr:02}, the problem is of more general nature. 
Already starting at
$O(\alpha_s^2)$, it is no longer true that a specific flavour current
$j_\mu=\bar{Q}\gamma_\mu Q$ contains only heavy quark production neither
that all heavy quark production originates from this current. As was 
shown in chapter 4 for the diagrams of fig. \ref{fig:4.b}, 
also light-light and heavy-light correlators contain
heavy quark production. Since only the total electromagnetic current,
including a sum over all flavours, is a physical observable, the single
flavour production is not uniquely defined. The crucial point is to
set up two identical quantities: the phenomenological and the 
theoretical side should be defined is such a way that they contain
the same contributions. The higher the order this may be a more
and more complicated task.


\section{Conclusions}

In this work we have obtained the following values for the
charm and bottom quark masses: 
\begin{eqnarray}
  \label{eq:10.a}
  M_c &=& 1.75 \pm 0.15 \ \mbox{GeV}\,,\qquad
	m_c=1.19 \pm 0.11 \ \mbox{GeV}\,, \nn\\ 
  M_b &=& 4.98 \pm 0.125 \ \mbox{GeV}\,,\qquad
	m_b=4.24 \pm 0.10 \ \mbox{GeV}\,.
\end{eqnarray}
As in the last section we evaluate the ${\rm \MSb}$-masses at their
own scale, $m_{c,b}=m_{c,b}(m_{c,b})$. Now 
we summarise the key features of this analysis.

In section 2 we have presented a rather complete setup for the 
quantum mechanical sum rules in the Coulomb potential. The correlator
contains poles below and a continuum
above threshold. The poles only start with a power of $O(\alpha^3)$, 
but exhibit an exponential behaviour \eqn{eq:2.m} in the Borel sum rules
or a sensitive power behaviour in the moment sum rules  \eqn{eq:2.u}.
Therefore the relative size of the poles depends strongly on either
the Borel parameter in the Borel sum rules or on $n$ and $\xi$ in
the moment sum rules. The analysis must be performed in a certain
sum rule window for $\tau$ or for $n$ and $\xi$ respectively to
guarantee a reliable theoretical calculation and sensitivity 
to the phenomenological parameters.

In the field theory case the expansion of the Green's function is
known up to NNLO in the framework of NRQCD. The Green's function
is directly evaluated at
$s_0=-4m^2\xi$ and in this way we avoid to sum up the energy levels
individually where the expansion is badly convergent. 
The spectral density can
be obtained from the imaginary part of the Green's function and the
pole contributions from the difference between the full and the
continuum result. The resulting moments depend on three scales:
$\mu_{soft}$, $\mu_{fac}$ and $\mu_{hard}$. 
In particular the dependence on  
$\mu_{soft}$ is relatively strong and presents the dominant
contribution to the error. 
Since the pole mass contains renormalon ambiguities we have also 
performed the analysis for the PS-mass which 
can be perturbatively related to the ${\rm \MSb}$-mass. 
Whereas the PS-mass by definition
depends on the separation scale which was used to subtract the
long-distance potential, this dependence cancels in the transition
to the ${\rm \MSb}$-mass to a large extend.
We have then included the perturbative contributions up to 
$O(\alpha_s^2)$. They are necessary to construct the spectral 
density above threshold for the full energy range and
to guarantee the stability of the
mass in a region of small $n$.

One of the great virtues of the method of QCD sum rules is the
analytic dependence on the theoretical and phenomenological
parameters. Thus we have investigated their importance and influence 
on the analysis. For the determination of the masses we have used
central values for all parameters.
These values were not motivated by any
optimisation or stability requirement, but only grounded on general
considerations.  Each was varied in a suitably large chosen window
for the error estimate.
Only the lower value of $\mu_{soft}$ was also limited by the
convergence of the nonrelativistic expansion. Finally, all errors
have been added quadratically.

We have set up the sum rules for an arbitrary evaluation point $\xi$.
With this parameter it is possible to shift the moments into a more
perturbative region for higher $\xi$ or into a region more sensitive
to the bound state energies and the mass for lower $\xi$.
We have used $\xi=0.5$ both for the charmonium and bottomium.
In fact, moving away from the threshold region and loosing
sensitivity on the mass, the scale dependence is even a bit larger
than at $\xi=0$. But the convergence of the theoretical expansions
is improved and the theoretical contributions more equally distributed 
among the different terms, in particular, the pole contributions
do not play such a dominant role. Thus we believe that we have reduced
the systematic  uncertainties in this sum rules which cannot be
accounted for by a variations of the scales.

We would like to emphasise a remarkable property of this analysis:  
Once a particular set of (central) values for 
$\mu_{soft}$, $\mu_{fac}$ and $\mu_{hard}$ is chosen, the
${\rm \MSb}$-masses remain very stable over a
large range of values for $n$, $\xi$ or $\mu_{sep}$.
In general, the ${\rm \MSb}$-bottom quark mass changes only by 
$\Delta m_b=\pm 30\ \mev$ from its central value for $4\le n\le 15$ 
and the charm quark mass by $\Delta m_c=\pm 15\ \mev$ for $3\le n\le 15
$.
Varying $\xi$ between $-0.25\le \xi\le 2$ the bottom quark mass changes by 
$\Delta m_b=^{+50}_{-40}\,\mev$ and the charm quark mass by 
$\Delta m_c=^{+50}_{-30}\,\mev$ for $-0.5\le \xi \le 2$. The variation of
$\mu_{sep}$ changes the PS-mass significantly since its definition
depends on $\mu_{sep}$. But in the relation to the  ${\rm \MSb}$-mass
this change is almost completely cancelled and the bottom quark mass changes
by $\Delta m_b=\pm 7 \ \mev$ for $1.0\ \gev \le \mu_{sep}\le 3.0\ \gev$
and the charm quark mass by $\Delta m_c=\pm 8 \ \mev$ for 
$0.8\ \gev \le \mu_{sep}\le 1.2\ \gev$.
These results are astonishing since the variation of these parameters
corresponds to largely different relative influence among the theoretical
contributions.
Thus we hope that we were able to set up a consistent framework
in which the physics of the relevant energy region, apart from the
remaining uncertainties in the nonrelativistic and perturbative expansions,
has been correctly described.

Let us finally summarise the achieved status.
In our analysis, several contributions seem to be under good control:
The perturbative expansion, as it has been incorporated in our analysis,
converges reasonably well. The condensates only give a
negligible contribution to the upsilon or charmonium.
On the phenomenological side, for the upsilon the first 
six resonances have been measured. For the non-resonant continuum
part, quark-hadron duality has been used. In the charmonium system, the
experimental situation has improved recently. Besides the first 
six resonances also the cross section between 2.0 GeV and 4.8 GeV
has been measured. Above this energy we again use quark-hadron duality.
Nevertheless, the most important contribution is already given
by the first two poles.

Of decisive importance for the determination of the masses is the
threshold behaviour. The method of QCD sum rules is a very powerful
tool to extract the masses since - by the choice of $n$ and $\xi$ -
it can react very sensitive to the threshold. Thus, large theoretical
uncertainties only lead to a relatively small shift in the masses.
The main uncertainties indeed come from the threshold expansion of NRQCD. 
The largest potential for an improvement
of the analysis lies in a further understanding of this energy region.
In particular, the knowledge of the Green's function at NNNLO could
help to reduce the error. 

The method of QCD sum rules is based on the assumptions of 
quark-hadron duality. With the development of NRQCD is has become 
clear that the pole contributions must be included in the theoretical
description for a correct comparison between the theoretical and
phenomenological part. The theoretical description is based on the operator 
product expansion and can be performed in a perturbative region where all
expansions converge well. However, this cannot be used
for an ever increasing precise determination of the mass. Since on the
phenomenological side the dependence on the mass originates mainly 
from the first bound states, the sensitivity on the mass decreases
in a similar way as the theoretical expansions improve.
In our analysis we have tried to balance these contributions 
by an appropriate choice of $\xi$ and $n$.
Without significant progress in the theoretical description
it seems that further substantial improvement will be difficult to achieve.

                 
\bigskip \noindent
{\bf Acknowledgements}

\noindent
It is a pleasure to thank Matthias Jamin for numerous
discussions, collaboration in parts of this work and reading
the manuscript.
I would like to thank Nora Brambilla, Antonio Pich, 
Jorge Portol{\'e}s, Pedro Ruiz and Antonio Vairo for helpful and interesting 
discussions and Antonio Pich for reading the manuscript. 
This work has been supported in part by TMR, EC
contract No. ERB FMRX-CT98-0169, by MCYT (Spain) under grant
FPA2001-3031, and by ERDF funds from the European Commission.
I thank the Deutsche Forschungsgemeinschaft for financial support.


\begin{appendix}


\section{Potential}

In NRQCD the Green's function obeys the Schr\"odinger
equation:
\begin{eqnarray}
  \label{ap:a.a}
  \lefteqn{\Bigg( -\frac{\Delta_x}{M}-\frac{\Delta_x^2}{4M^3}+V_C(x)
  +\frac{\alpha_s}{4\pi}\Delta_1 V(x)
	+\frac{\alpha_s^2}{16\pi^2}\Delta_2 V(x)} \nn\\
  && +\Delta_{NA}V(x) 
  +\Delta_{BF}V(x)+\frac{k^2}{M}\Bigg)
  G({\bf x},{\bf y},k)=\delta^{(3)}({\bf x}-{\bf y}) \,.
\end{eqnarray}
$V_C(x)$ is the Coulomb potential,
$\Delta_{NA}V(x)$ the nonabelian part of the quark-antiquark potential,
$\Delta_{BF}V(x)$ the Breit-Fermi potential and the terms
$\Delta_1 V(x)$ and $\Delta_2 V(x)$ contain the first and second
order perturbative correction to the Coulomb potential.
The explicit expressions read:
\begin{eqnarray}
  \label{ap:a.b}
  V_C(x)&=& -\frac{C_F\alpha_s}{x} \,,\qquad x=|{\bf x}|\,,\nn\\
  \Delta_{NA}V(x) &=& -C_A C_F \frac{\alpha_s^2}{2Mx^2} \,,\nn\\
  \Delta_{BF}V(x) &=& \frac{C_F \alpha_s \pi}{M^2}\delta^{(3)}({\bf x})
  -\frac{C_F \alpha_s}{2M^2 x}\left({\bf p}^2+\frac{1}{x^2}{\bf x}({\bf x p})
      {\bf p}\right)+\frac{3 C_F \alpha_s}{2M^2x^3}{\bf S L} \nn\\
  && -\frac{C_F \alpha_s}{2M^2}\left(\frac{{\bf S}^2}{x^3}-
    \frac{3({\bf S x})^2}{x^5} -\frac{4\pi}{3}
    \left(2{\bf S}^2-3\right)\delta^{(3)}({\bf x}) \right) \,,\nn\\
  \Delta_1 V(x) &=& V_C(x) \left(
    a_1+2 b_0 \gamma_E+2 b_0 \ln (x\mu)\right)\,,\nn\\
  \Delta_2 V(x) &=& V_C(x)\Big(
    a_2+b_0^2\left(\pi^2/3+4 \gamma_E^2\right)+2 \gamma_E(b_1+2 b_0 a_1)\nn\\
    && +\left(2 b_1+4 b_0 a_1+8b_0^2 \gamma_E\right) \ln (x\mu) +
    4 b_0^2 \ln^2 (x\mu) \Big) \,,
\end{eqnarray}
with the constants
\begin{eqnarray}
  \label{ap:a.c}
  b_0 &=& 11-\frac{2n_f}{3} \,, \nn\\
  b_1 &=& \frac{34}{3}C_A^2-\frac{20}{3}C_A T n_f - 4 C_F T n_f
  =102-12.67 n_f\,,\nn\\
  a_1 &=& \frac{31}{9} C_A-\frac{20}{9}T n_f 
  = 10.33-1.11 n_f\,,\nn\\
  a_2 &=& \left(\frac{4343}{162}+4\pi^2-\frac{\pi^4}{4}+
	\frac{22}{3}\zeta(3)\right)
  C_A^2 -\left(\frac{1798}{81}+\frac{56}{3}\zeta(3)\right)C_A T n_f \nn\\
  &&-\left(\frac{55}{3}-16 \zeta(3)\right)C_F T n_f + 
	\left(\frac{20}{9}T n_f\right)^2\nn\\
  &=& 456.75-66.35 n_f+1.23 n_f^2 \,,
\end{eqnarray}
where $C_F=4/3$, $C_A=3$ and $T=1/2$.
The coefficient $a_2$ was first calculated in \cite{p:97}
and later corrected to the above given value \cite{s:99}.


\section{Potential-Subtracted mass}

The PS-mass is defined by \cite{b:98}
\begin{eqnarray}
  \label{ap:b.a}
  \delta m(\mu_{sep}) &=& -\frac{1}{2}\int
	\limits_{|{\bf q}|<\mu_{sep}}\frac{d^3 q}{(2\pi)^3}\,V(q) \,,\nn\\
  m_{PS}(\mu_{sep}) &=& M-\delta  m(\mu_{sep})\,.
\end{eqnarray}
The subtracted potential $V(r,\mu_{sep})$ is related to the potential
in momentum space  $V(q)$ and $\delta m(\mu_{sep})$:
\begin{eqnarray}
  \label{ap:b.b}
  V(q) &=& - \frac{4\pi C_F \alpha_s({\bf q})}{{\bf q}^2} 
    \left(1+a_1 \frac{\alpha_s ({\bf q})}{4\pi}+a_2 
	\left(\frac{\alpha_s({\bf q})}{4\pi}
        \right)^2\right) \,,\nn\\
  V(r) &=& \int \frac{d^3 q}{(2\pi)^3}\,e^{i{\bf qr}}\, V(q)\,,\nn\\
  V(r,\mu_{sep}) &=& V(r)+2\delta m(\mu_{sep}) \,,
\end{eqnarray}
with $a_1$ and $a_2$ as in eq. \eqn{ap:a.c}. Performing the
Fourier transformation one obtains the relation between
PS- and pole mass:
\begin{eqnarray}
  \label{ap:b.c}
  m_{PS}(\mu_{sep}) &=& M \Big[1+a(\mu)r^{(1)}_{PS}(\mu_{sep})+
  a^2(\mu)r^{(2)}_{PS}(\mu,\mu_{sep})
  + a^3(\mu)r^{(3)}_{PS}(\mu,\mu_{sep})\Big] \,,\nn\\
  r^{(1)}_{PS}(\mu_{sep}) &=& -C_F\frac{\mu_{sep}}{M} \,, \qquad
  r^{(2)}_{PS}(\mu,\mu_{sep}) = -C_F\frac{\mu_{sep}}{M}\frac{w_1(\mu,\mu_{sep})}{4} \,,\nn\\
  r^{(3)}_{PS}(\mu,\mu_{sep}) &=& -C_F\frac{\mu_{sep}}{M}\frac{w_2(\mu,\mu_{sep})}{16} \,,
\end{eqnarray}
with $a=\alpha_s/\pi$ and the functions
\begin{eqnarray}
  \label{ap:b.d}
  w_1(\mu,\mu_{sep}) &=& a_1-b_0\left(\ln 
	\frac{\mu_{sep}^2}{\mu^2}-2\right) \,,\nn\\
  w_2(\mu,\mu_{sep}) &=& a_2-\left(2a_1 b_0+b_1\right)
	\left(\ln \frac{\mu_{sep}^2}{\mu^2}-2\right)
  +b_0^2\left(\ln^2 \frac{\mu_{sep}^2}{\mu^2}
	-4\ln \frac{\mu_{sep}^2}{\mu^2}+8\right) \,,\nn\\
\end{eqnarray}
with $b_0$ and $b_1$ from eq. \eqn{ap:a.c}.
Using the three-loop result between pole and ${\rm \MSb}$-mass
one can relate the PS- and ${\rm \MSb}$-mass:
\begin{eqnarray}
  \label{ap:b.e}
  m_{PS}(\mu_{sep}) &=& m \Bigg[ 1+ a(m)
  \left(k_1-C_F\frac{\mu_{sep}}{m}\right)
  +a^2(m) \left(k_2 
  -C_F\frac{\mu_{sep}}{m}\frac{w_1(m,\mu_{sep})}{4}
    \right) \nn\\
    &&+ a^3(m)\left(k_3-C_F\frac{\mu_{sep}}{m}\frac{w_2(m,\mu_{sep})}{16}\right)      \Bigg] \,,\nn\\
  k_1 &=& C_F \,, \qquad k_2 = 13.443 - 1.041 n_f \,,\nn\\
  k_3 &=& 190.595 -26.655 n_f + 0.653 n_f^2 \,,
\end{eqnarray}
where $m=m_{\MSb}(m_{\MSb})$ is the ${\rm \MSb}$-mass evaluated at its
own scale.  

\end{appendix}

\end{document}